\documentclass[reprint,
superscriptaddress,
citeautoscript,
amsmath,amssymb,
aps,
prb
]{revtex4-1}

\usepackage[pdftex]{graphicx}
\usepackage{dcolumn}
\usepackage{bm}
\usepackage{hyperref}

\begin{document}

\preprint{APS/123-QED}

\title{Magnetic phase diagram of low-doped La$_{2-x}$Sr$_x$CuO$_4$ thin films studied by\\ low-energy muon-spin rotation}

\author{E. Stilp}
 \affiliation{Laboratory for Muon Spin Spectroscopy, Paul Scherrer Institut, CH-5232 Villigen PSI, Switzerland}
 \affiliation{Physik-Institut der Universit\"at Z\"urich, Winterthurerstrasse 190, CH-8057 Z\"urich, Switzerland}
 
\author{A. Suter}
 \affiliation{Laboratory for Muon Spin Spectroscopy, Paul Scherrer Institut, CH-5232 Villigen PSI, Switzerland}

\author{T. Prokscha}
 \affiliation{Laboratory for Muon Spin Spectroscopy, Paul Scherrer Institut, CH-5232 Villigen PSI, Switzerland}

\author{E. Morenzoni}
 \affiliation{Laboratory for Muon Spin Spectroscopy, Paul Scherrer Institut, CH-5232 Villigen PSI, Switzerland}

\author{H. Keller}
 \affiliation{Physik-Institut der Universit\"at Z\"urich, Winterthurerstrasse 190, CH-8057 Z\"urich, Switzerland}

\author{B.~M.~Wojek}
 \altaffiliation{Present address: KTH Royal Institute of Technology, ICT Materials Physics, Electrum 229, 164 40 Kista, Sweden}
 \affiliation{Laboratory for Muon Spin Spectroscopy, Paul Scherrer Institut, CH-5232 Villigen PSI, Switzerland}
 \affiliation{Physik-Institut der Universit\"at Z\"urich, Winterthurerstrasse 190, CH-8057 Z\"urich, Switzerland}
 
\author{H. Luetkens}
 \affiliation{Laboratory for Muon Spin Spectroscopy, Paul Scherrer Institut, CH-5232 Villigen PSI, Switzerland}
 
\author{A. Gozar}
 \affiliation{Brookhaven National Laboratory, Upton, New York 11973-5000, USA}

\author{G. Logvenov}
 \altaffiliation{Present address: Max Planck Institute for Solid State Research, Heisenbergstrasse 1, D-70569 Stuttgart, Germany}
 \affiliation{Brookhaven National Laboratory, Upton, New York 11973-5000, USA}

\author{I. Bo\v{z}ovi\'{c}}
 \affiliation{Brookhaven National Laboratory, Upton, New York 11973-5000, USA}

\date{\today}

\begin{abstract}

The magnetic phase diagram of La$_{2-x}$Sr$_x$CuO$_4$ thin-films grown on single-crystal LaSrAlO$_4$ substrates has been determined by low-energy muon-spin rotation. The diagram shows the same features as the one of bulk La$_{2-x}$Sr$_x$CuO$_4$, but the transition temperatures between distinct magnetic states are significantly different. In the antiferromagnetic phase the N\'{e}el temperature $T_{\rm N}$ is strongly reduced, and no hole spin freezing is observed at low temperatures. In the disordered magnetic phase ($x\gtrsim0.02$) the transition temperature to the cluster spin-glass state $T_{\rm g}$ is enhanced. Possible reasons for the pronounced differences between the magnetic phase diagrams of thin-film and bulk samples are discussed.
\end{abstract}

\maketitle

\section{Introduction}
\label{intor}

\begin{figure}[b!]
	\centering
		\includegraphics[width=\columnwidth]{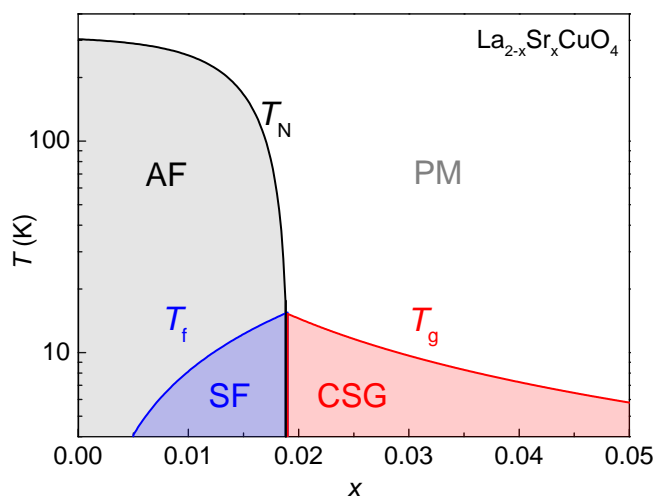}
	\caption{(color online) Schematic magnetic phase diagram of bulk LSCO. The N\'{e}el temperature $T_{\rm N}$, the spin freezing temperature of doped holes $T_{\rm f}$, and the glass transition temperature $T_{\rm g}$ are shown in dependence of the Sr content $x$. AF: antiferromagnetic; SF: spin freezing; CSG: cluster spin-glass; PM: paramagnetic.}
	\label{fig:Phasediagram1}
\end{figure}

The bulk magnetic phase diagram of the cuprates, especially La$_{2-x}$Sr$_{x}$CuO$_{4}$ (LSCO), has been extensively studied in the past two decades~\cite{Kastner98,Dagotto94,Sato2000,Rigamonti06}. At lowest Sr contents ($x \lesssim 0.02$) bulk LSCO is an antiferromagnetic (AF) charge-transfer insulator. Long-range 3D AF order appears below the N\'{e}el temperature $T_{\rm N} \simeq 300$\,K in the parent compound La$_2$CuO$_4$ (LCO)~\cite{Kastner98,Cho93,Chou93,Borsa95,Niedermayer98}. It results from the ordering of spin-$1/2$ Cu$^{2+}$ moments due to super-exchange with the in-plane exchange coupling constant $J/k_{\rm B}\simeq 1500$\,K. LCO is considered as a model system of a spin-$1/2$ quasi-2D Heisenberg antiferromagnet on a square lattice. The in-plane magnetic properties are well described by the so-called renormalized classical regime as derived by Chakravarty \textit{et al.}~\cite{Chakravarty88} and experimentally verified~\cite{Keimer92,Matsumura94}. The 3D order is established predominantly by the weak out-of-plane exchange coupling $J' \simeq 10^{-5}\, J$~\cite{Keimer92}. At a nominal Sr content of $x\simeq 0.02$ the N\'{e}el temperature decreases to zero as shown in the schematic phase diagram in Fig.~\ref{fig:Phasediagram1}. Within the AF state, charge localization of the doped holes is observed below the freezing temperature $T_{\rm f}$  [``spin freezing'' (SF)] which depends linearly on the doping level, as $T_{\rm f}=815$\,K$\cdot x$ (Ref.~\onlinecite{Chou93}).

For nominal Sr contents $x\gtrsim 0.02$ short-range AF correlations within the CuO$_2$ planes persist. A complicated interplay between the doped holes in the CuO$_2$ planes and the remaining AF correlations leads to a yet not well understood electronic state with a pseudogap in the excitation spectrum~\cite{Ginsberg89}. In this doping region, below the freezing temperature $T_{\rm g}$ spontaneous zero-field precession is observed in muon-spin rotation ($\mu$SR) studies~\cite{Chou93,Borsa95}. This is often referred to as the ``cluster spin-glass'' (CSG) phase. The ``glass'' transition temperature $T_{\rm g}$ decreases as $1/x$ (Ref.~\onlinecite{Chou93}) and is also detected within the superconducting phase that starts at $x\simeq 0.05$.

Thin films open the door to new physical properties and phenomena, since electronic or magnetic properties of thin-film structures can be very different from those of the single constituents as found in bulk samples. Phenomena driven by various couplings and dimensional effects may appear. For instance, the proximity between different orders can be studied in multi-layer systems and superlattices. In different cuprate heterostructures a giant proximity effect has been found~\cite{Bozovic04,Morenzoni11,Wojek12}, where low doped cuprates sandwiched between superconducting layers, can transmit supercurrent or exhibit a Meissner effect over surprisingly large distances at temperatures where these layers are intrinsically in the normal state. It is usually assumed that the magnetic layers in thin film systems behave as in the bulk material. Yet, this was never systematically studied. As a local magnetic probe of thin films low-energy muon-spin rotation (LE-$\mu$SR) is well suited to address this question~\cite{Prokscha08}. Previous studies by this technique of canonical spin glasses~\cite{Morenzoni08}, metal-insulator LSCO superlattices~\cite{Suter2011}, and nickel-oxide superlattices~\cite{Boris2011} showed that dimensional effects might strongly influence the magnetic ground state and its excitations. In addition, the mismatch between the lattice constants of the thin film material and of the substrate leads to biaxial positive or negative strain in the film. For example, LSCO grown on single-crystal LaSrAlO$_4$ (LSAO) substrates are under compressive strain, whereas on single-crystal SrTiO$_3$ (STO) substrates they are under tensile strain. Epitaxial strain leads to significant changes in the lattice constants of the films (contraction or expansion), which in turn affects the superconducting transition temperature~\cite{Sato97,Locquet98,Si99,Bozovic02,Si01} as well as the electronic band structure~\cite{Abrecht03}. Since both the substrates and the cuprates are essentially ionic crystals, apart from the above ``geometric'' effect (``Poisson strain'')~\cite{Bozovic02}, there is an additional effect due to un-screened, long-range Coulomb interactions (``Madelung strain'')~\cite{Butko09}, which manifests itself as a significant change in the unit cell volume. Finally, in a preliminary study by Suter \textit{et al.}~\cite{Suter04} changes were also observed in $T_{\rm N}$ of LCO, depending on the choice of the substrate. Therefore, the question arises how epitaxial strain, potential strain release, and the substrate in general affect the magnetic properties of LSCO thin films.

Here we present a study on the magnetic phase diagram of LSCO thin films in the low-doping regime (thickness $\simeq 53$\,nm, $0.00 \leq x \leq 0.06$) grown on LSAO. In Sec.~\ref{sec:exdet} the experimental details are given. Secs.~\ref{sec:AF} and B present the LE-$\mu$SR results of the AF and CSG phase, respectively. This technique allows stopping muons in matter at different depths in the nanometer range~\cite{Morenzoni04}, and is therefore well suited to investigate magnetic thin-film samples on a microscopic scale. In Sec.~\ref{sec:Discussion} the differences in the magnetic phase diagrams as obtained for bulk and thin-film samples are discussed, followed by the summary and conclusions in Sec.~\ref{sec:conclusions}.

\section{Experimental Details}
\label{sec:exdet}

\begin{figure}[b]
	\centering
		\includegraphics[width=\columnwidth]{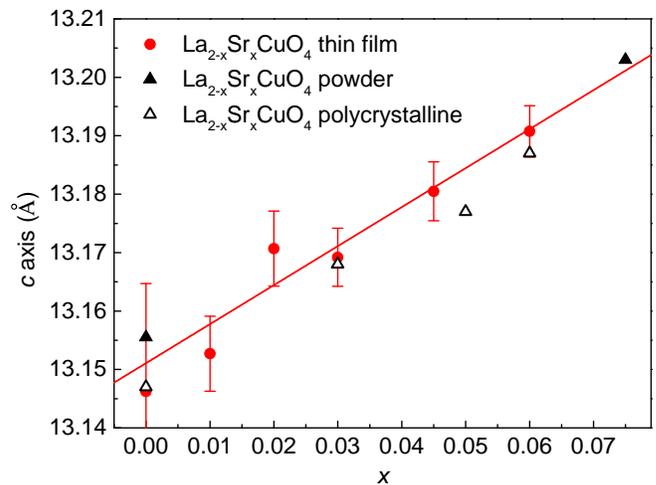}
	\caption{(color online) The crystallographic $c$ axis lattice constant as a function of the nominal Sr content $x$ in $53$\,nm thick LSCO films deposited on single-crystal LSAO substrates, as determined by x-ray diffraction (red circles) and in LSCO bulk samples (black open~\cite{Lampakis00} and filled~\cite{Radaelli94} triangles). The red solid line is a linear fit to the thin film data.}
	\label{fig:Xray}
\end{figure}

\begin{figure}[t]
	\centering
		\includegraphics[width=\columnwidth]{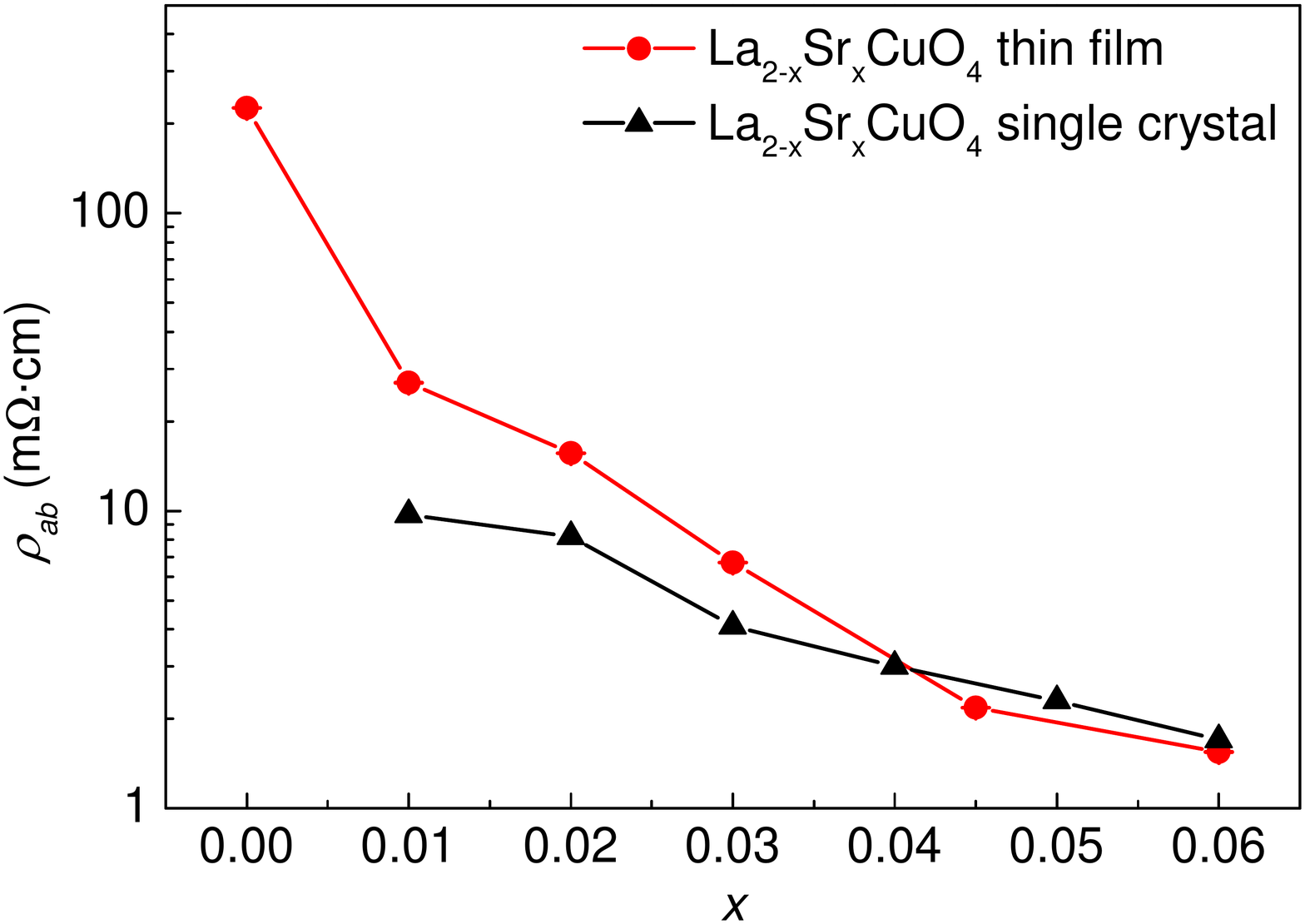}
	\caption{(color online) Resistivity $\rho_{ab}$ at $T=300$\,K versus the nominal Sr content $x$ of MBE-grown LSCO thin films (red circles) and LSCO single crystals (black triangles, from Ref.~\onlinecite{Ando04}). The lines are guides to the eye.}
	\label{fig:RversusX}
\end{figure}

The La$_{2-x}$Sr$_x$CuO$_4$ films studied here were synthesized using molecular-beam epitaxy (MBE) at the Brookhaven National Laboratory. We used single crystal LSAO substrates, $10 \times 10 \times 1$\,mm$^3$ in size and poilished with the surface perpendicular to the [001] crystal axis. The typical film thickness was $53$\,nm. Further information about the growing process has been published elsewhere~\cite{Bozovic01}. Here, we have investigated thin films with $x=0.00$, $0.01$, $0.02$, $0.03$, $0.045$, and $0.06$. The doping level was controlled during the deposition by using well-calibrated MBE sources; the rates were monitored and controlled in real time using a custom-built 16-channel atomic absorption spectroscopy system~\cite{Bozovic01}. The film growth and quality was monitored in real time using reflection high-energy electron diffraction (RHEED), and checked subsequently by atomic force microscopy as well as by resistivity, susceptibility, and x-ray diffraction measurements. The $c$ axis lattice parameters of the samples were extracted from $\theta$-2$\theta$-scans (Fig.~\ref{fig:Xray}). They show a linear behavior as function of the nominal Sr content $x$, and agree well with the bulk data of polycrystalline samples~\cite{Lampakis00} and powder samples~\cite{Radaelli94}. A comparison of the room-temperature resistivity for LSCO single crystals~\cite{Ando04} and the investigated thin films as function of $x$ is depicted in Fig.~\ref{fig:RversusX}. The resistivity at $300$\,K shows the expected decrease with increasing Sr content $x$. The thin-film resistivity data are comparable to the single crystal data.

To study the magnetic phase diagram of thin-film LSCO, LE-$\mu$SR experiments were performed at the muE4 beamline at the Paul Scherrer Institute (PSI, Switzerland)~\cite{Prokscha08}. In a $\mu$SR experiment positively charged muons $\mu^{+}$ with $\sim 100\,\%$ spin polarization are implanted in the sample where they thermalize within a few picoseconds without noticeable loss of polarization. Because of interactions of the $\mu^{+}$ spins with internal local magnetic fields $B_{\rm loc}$ the magnetic moments of the $\mu^{+}$ precess with the Larmor frequency $\omega_{\rm L} = \gamma_{\mu} B_{\rm loc}$ ($\gamma_{\mu} = 2\pi \times 135.54$\,MHz/T) in the sample until they decay with a mean lifetime of $\tau_{\mu} = 2.197$\,$\mu$s into neutrinos ($\bar{\nu}_{\mu}$, $\nu_{\rm e}$) and positrons (${\rm e}^{+}$):
\begin{equation*}
\mu^{+} \rightarrow {\rm e}^{+} + \bar{\nu}_{\mu} + \nu_{\rm e}.
\end{equation*}

\begin{figure}[t]
	\centering
		\includegraphics[width=\columnwidth]{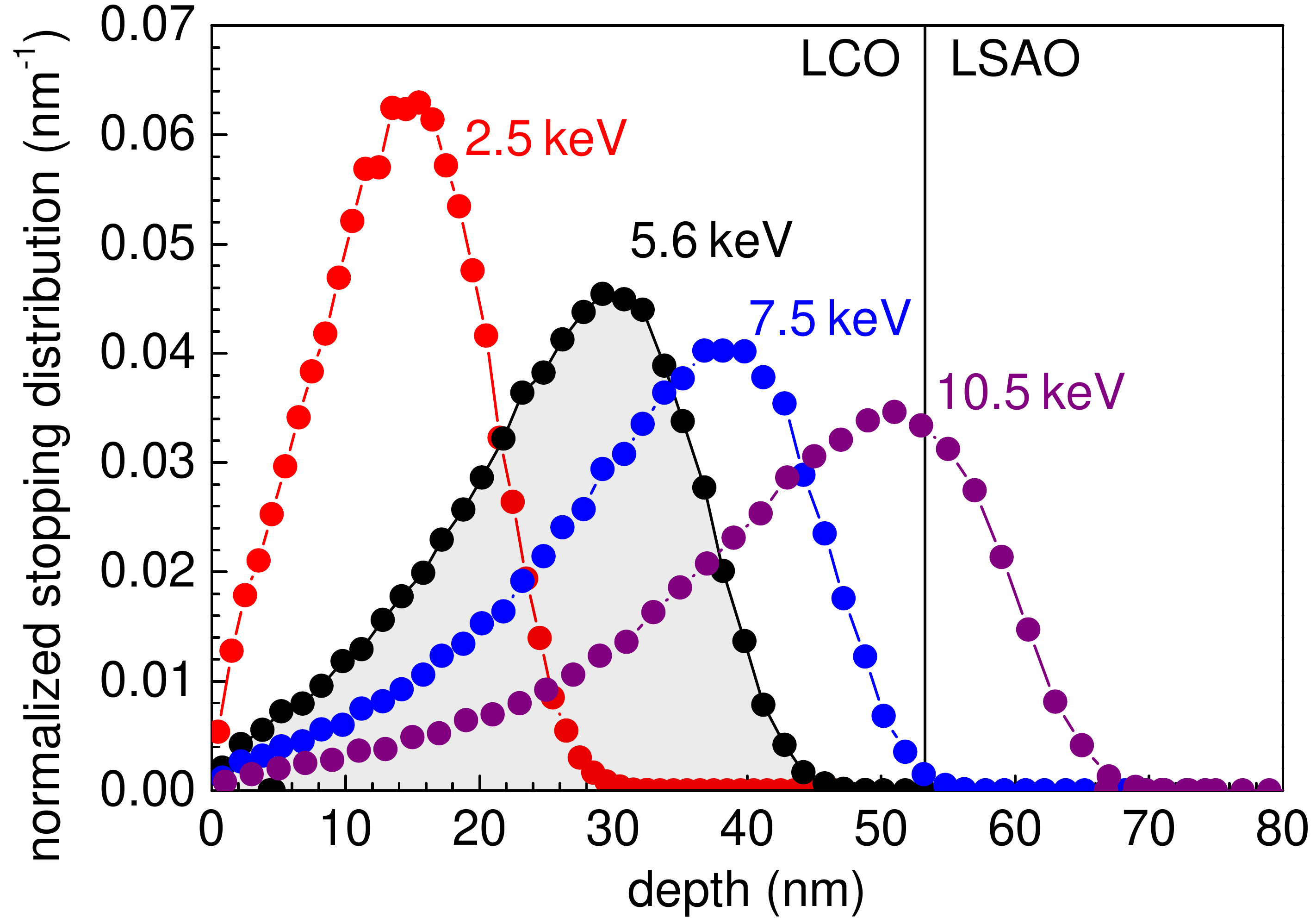}
	\caption{(color online) The normalized stopping distribution of muons with different implantation energies (numbers given in the figure) of a $53$\,nm thick LCO film deposited on a single-crystal LSAO substrate calculated using TRIM.SP~\cite{Eckstein91}. The lines are guides to the eye.}	
	\label{fig:trimsp}
\end{figure}

The emission probability for the positron along the muon spin direction is enhanced due to the parity-violating muon decay. Measuring the time difference $t=t_{\rm e} - t_{\rm s}$ between the implantation time $t_{\rm s}$ of the $\mu^{+}$ and its decay time $t_{\rm e}$, detected via the decay positron (for $\sim 5 \times 10^6$ $\mu^{+}$) allows one to determine the temporal evolution of the muon-spin polarization $P(t)$ (time ensemble average) via the positron count rate $N(t)$:
\begin{equation}
\centering
N(t) = N_0 \, {\rm e}^{-t/\tau_\mu} \, \left[1 + A \, P(t)\right] + N_{\rm Bkg},
\end{equation}
were $N_0$ gives the scale of the counted positrons, $N_{\rm Bkg}$ is a time-independent background of uncorrelated events, and $A$ is the observable decay asymmetry. The latter is a function of the positron energy and the solid angle of the positron detectors. In our experimental setup $A$ $\simeq 0.25=A_{\rm max}$. The exponential function describes the radioactive muon decay. From the measured $P(t)$ one can extract the local magnetic fields, field distributions, and field fluctuations present in the sample~\cite{Yaouanc11}. In bulk $\mu$SR experiments $\mu^{+}$ with an energy of $\sim 4.1$\,MeV are used, which originate from the positively charged pion decay at rest at the surface of the muon production target (``surface muons''). In this case the mean stopping depth in condensed matter is of the order of $\sim 100 \, \mu$m. To investigate thin films LE-$\mu$SR makes use of epithermal muons ($\sim 15$\,eV). They are created by moderating surface muons~\cite{Morenzoni04,Harshman87}. After reacceleration, the final muon implantation energy is controlled by applying a voltage to the sample. By tuning the energy between $1$\,keV and $30$\,keV, mean depths between a few and a few hundred nanometers can be chosen. The normalized stopping distribution of $\mu^{+}$ in a LCO film deposited on a LSAO substrate for different implantation energies is depicted in Fig.~\ref{fig:trimsp}.

For each Sr content, we used a mosaic of four thin-film samples, each with lateral dimensions of $1\times1$\,cm$^2$, glued onto a silver-coated aluminum plate with silver paint. To reach temperatures in the range $3$\,K to $300$\,K a cold-finger cryostat was used. The experiments were performed in ultra-high vacuum at a pressure of about $10^{-9}$\,mbar. The data presented here were all obtained with a muon implantation energy $E_{\rm impl.} = 5.6$\,keV. For this energy Monte Carlo simulations performed using TRIM.SP~\cite{Eckstein91} yield a mean implantation depth of about $30$\,nm which is optimal for these films (Fig.~\ref{fig:trimsp}). In order to check the homogeneity of the films across their thickness $\mu$SR time spectra $N(t)$ for different values of $E_{\rm impl.}$ were measured showing no differences. LE-$\mu$SR measurements were performed in zero magnetic field (ZF) to determine the internal magnetic fields at the muon stopping site, which are related to the staggered magnetization, as well as in weak transverse magnetic fields (wTF) in the range of $2.8$\,mT to $9.8$\,mT to obtain the magnetic transition temperatures $T_{\rm N}$, $T_{\rm f}$, $T_{\rm g}$, and the magnetic volume fractions $f$.

\section{Results and Discussion}
\label{results}

\subsection{Antiferromagnetic regime}
\label{sec:AF}

\begin{figure}[t]
	\centering
		\includegraphics[width=\columnwidth]{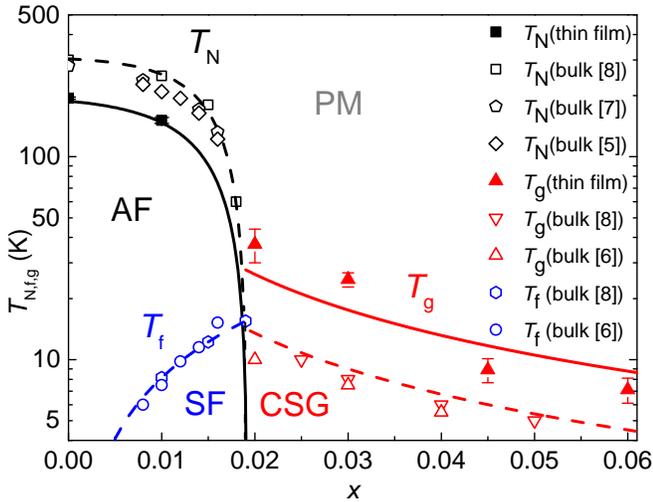}
	\caption{(color online) The magnetic phase diagram of LSCO for thin-films (solid lines) and bulk samples (dashed lines). The N\'{e}el temperature $T_{\rm N}$, the freezing temperature $T_{\rm f}$, and the glass transition temperature $T_{\rm g}$ are shown as a function of the nominal Sr content $x$. The black lines are guides to the eye. The blue and red lines follow the relations $T_{\rm f} \propto x$ and $T_{\rm g} \propto 1/x$, respectively. AF: antiferromagnetic; SF: spin freezing; CSG: cluster spin-glass; PM: paramagnetic.}
	\label{fig:Phasediagram}
\end{figure}

We first investigated thin-film samples with $x=0.00$ and $x=0.01$ in the AF regime of the phase diagram. From temperature scans in a weak magnetic field the N\'{e}el temperatures were determined to be $T_{\rm N}^{x=0.00}=195(3)$\,K and $T_{\rm N}^{x=0.01}=151(5)$\,K (Fig.~\ref{fig:Phasediagram}). These values are much lower compared to bulk values, as will be discussed later in detail.

\begin{figure}[b]
	\centering
		\includegraphics[width=1\columnwidth]{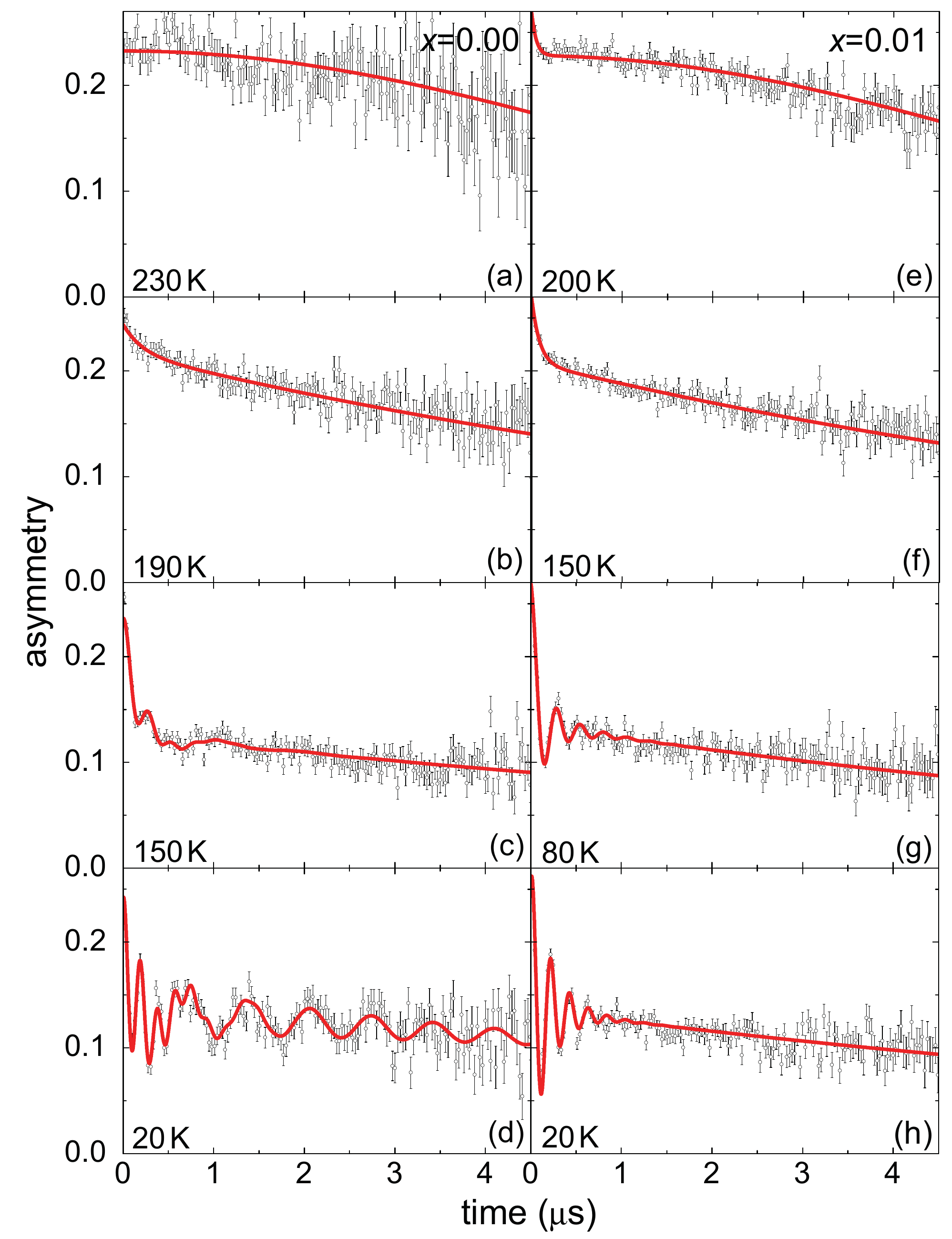}
	\caption{(color online) The ZF asymmetry time spectra of LSCO thin films with $x=0.00$ (left panels) and $x=0.01$ (right panels) at different temperatures for $E_{\rm impl.}=5.6$\,keV. The solid red lines are fits to the data done with musrfit~\cite{Suter12}. See text for more details.}
	\label{fig:LCO}
\end{figure}

In the paramagnetic (PM) state ($T \gg T_{\rm N}$) the asymmetry time spectra in ZF, $A\,P^{\rm PM}_{\rm ZF}(t)$, are well described by a Gaussian Kubo-Toyabe function [Fig.~\ref{fig:LCO}\,(a)], corresponding to a 3D Gaussian field distribution of dense randomly oriented static magnetic moments:
\begin{equation}
\centering
A\,P^{\rm PM}_{\rm ZF}\left(t\right) = A \ \left[ \frac{1}{3} + \frac{2}{3} \left(1-\sigma^2t^2\right) \, {\rm e}^{-\frac{1}{2} \sigma^2 t^2}\right],
\end{equation}
where $A$ is the decay asymmetry and $\sigma$ the depolarization rate. This is expected since only the nuclear moments of La and Cu contribute to $A\,P^{\rm PM}_{\rm ZF}(t)$. The PM fluctuation rate of the electronic Cu moments is too high to have an observable influence on the ZF spectra. The nuclear moments, however, are static on $\mu$SR time scales. In all ZF fits a temperature-independent constant background asymmetry $A_{\rm Bkg}=0.17(3) \cdot A_{\rm max}$ was taken into account, which originates from the muons stopping in the silver coating of the sample plate. Only about $80\,\%$ to $85\,\%$ of the muons stop in the sample. Close to the magnetic transition ($T \gtrsim T_{\rm N}$) the time spectrum $A\,P(t)$ changes first to a combination of a Gaussian Kubo-Toyabe function with an exponential decay [Fig.~\ref{fig:LCO}\,(e)] and then to a superposition of exponential decay functions [Fig.~\ref{fig:LCO}\,(b)~and\,(f)]. At these temperatures the electronic fluctuations slow down, giving rise to a stronger depolarization of the muons. This behavior is also observed in the PM phase of bulk samples~\cite{Borsa95}.

\begin{figure*}[t]
\centering
  \begin{minipage}{4 cm}
    \includegraphics[width=0.8\columnwidth]{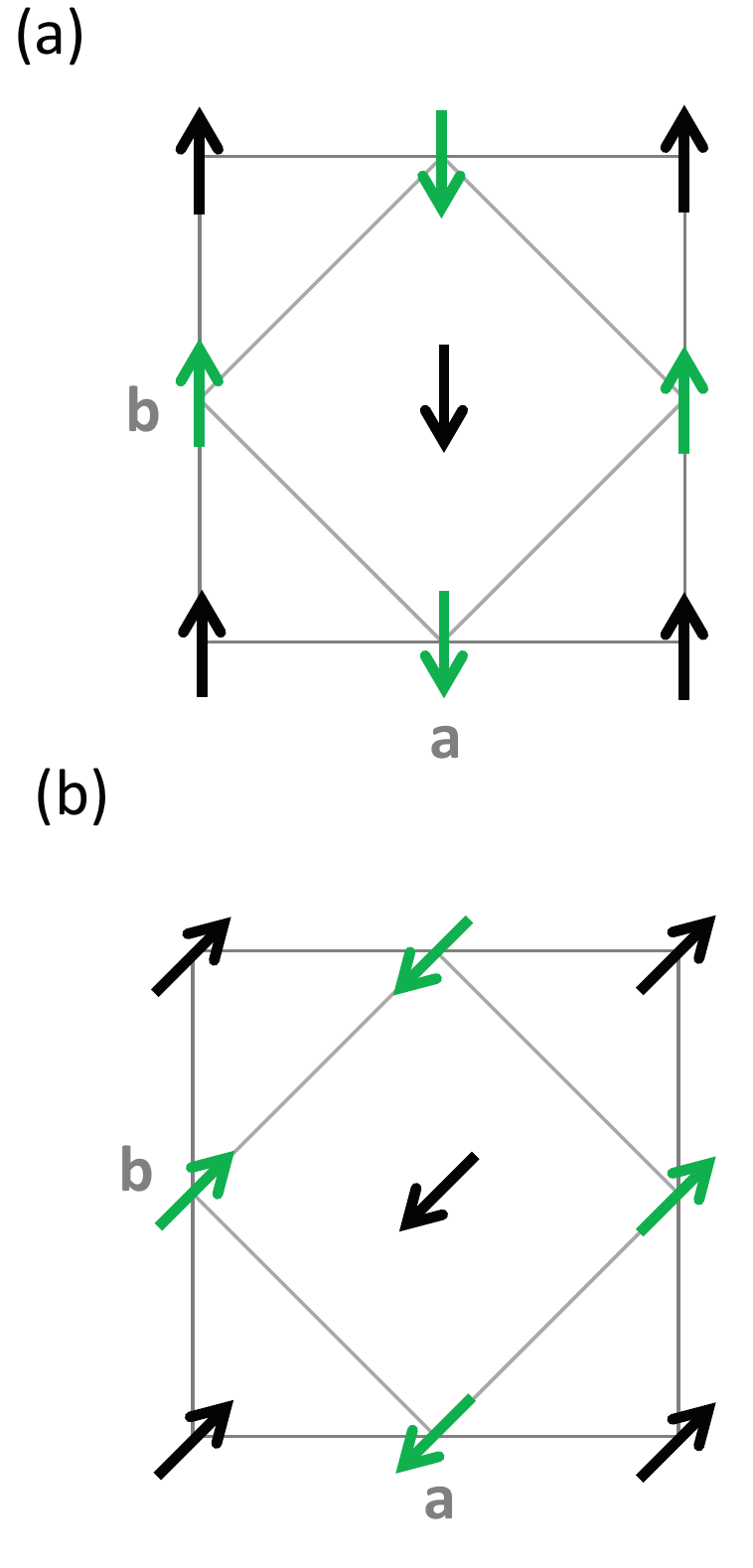}  
  \end{minipage}
  \begin{minipage}{9.9 cm}
    \includegraphics[width=1\columnwidth]{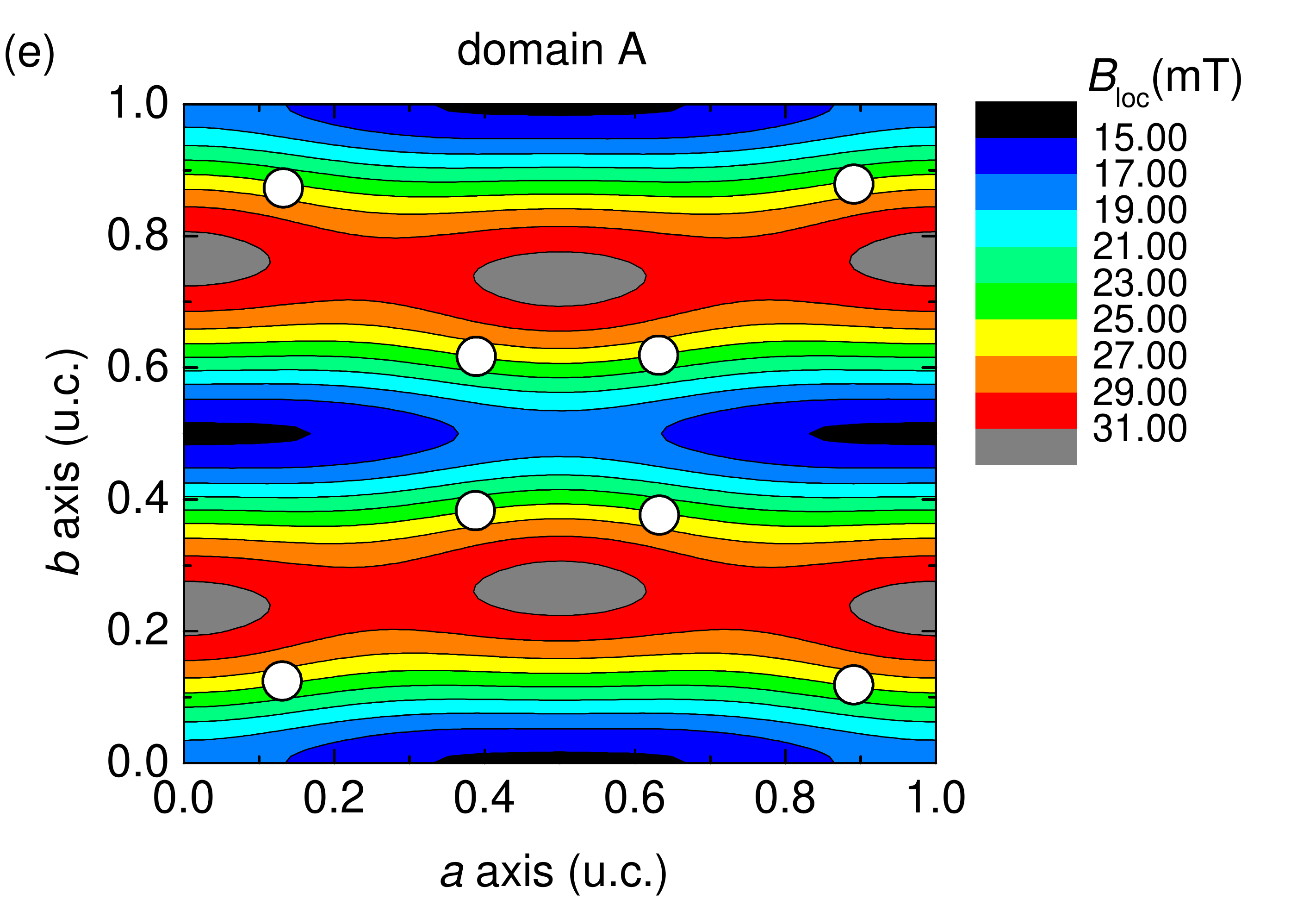}  
  \end{minipage}
  \begin{minipage}{4 cm}
    \includegraphics[width=0.8\columnwidth]{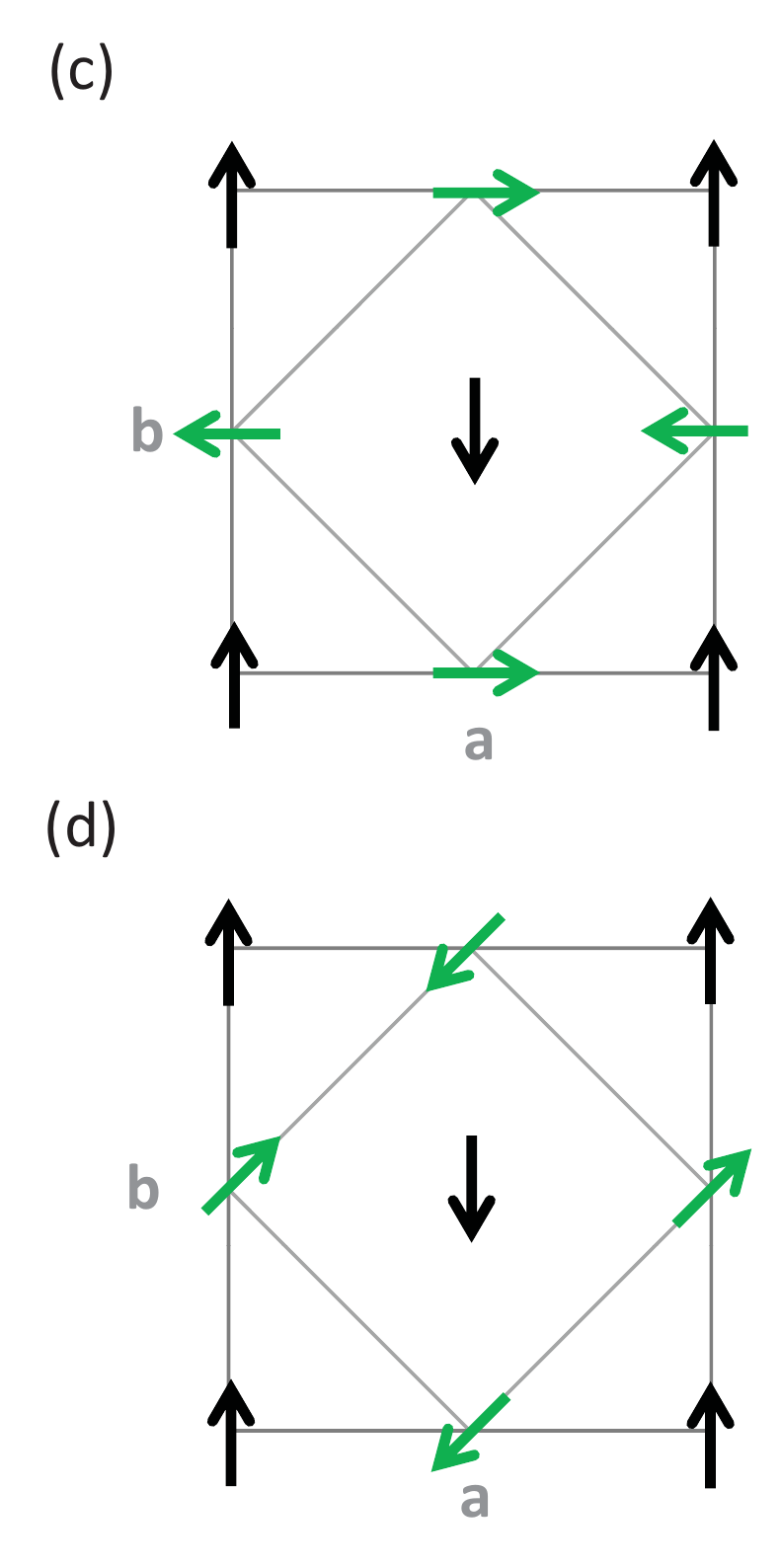}
  \end{minipage} 
  \begin{minipage}{9.9 cm}
    \includegraphics[width=1\columnwidth]{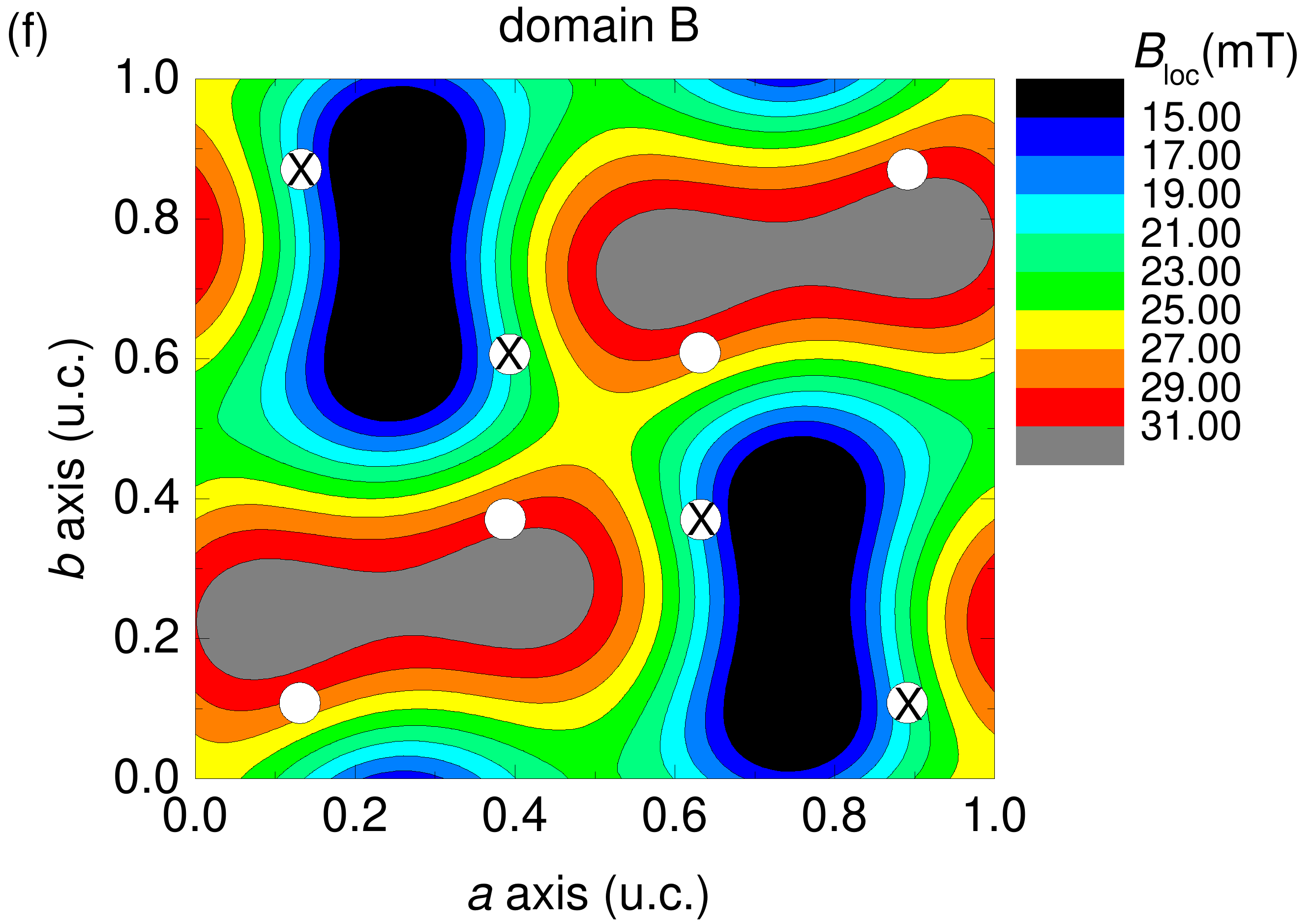}  
  \end{minipage}
  \caption{(color online) (a)\,-\,(d) show different arrangements of the Cu electronic magnetic moments as viewed along the $c$ axis of the orthorhombic unit cell of LSCO ($a=5.3568$\,\AA{}, $b=5.4058$\,\AA{}, $c=13.1432$\,\AA{})~\cite{Reehuis06}. Configuration (a) leads to one local magnetic field $B_{\rm loc}$ at the muon stopping site (``domain A''), whereas configurations (b)\,-\,(d) exhibit two $B_{\rm loc}$ (``domain B''). The black and green arrows in (a)\,-\,(d) correspond to moments in adjacent CuO$_2$ layers ($z=0$ and $z=c/2$, respectively). (e) and (f) show the resulting magnetic field maps for the spin arrangements (a) and (b) for the plane $z=0.2128\,c$ based on dipole field distribution calculations. The circles mark the muon stopping sites within the unit cell were $B_{\rm loc,1}$ (open circles) or $B_{\rm loc,2}$ (circles with crosses) is present.}
  \label{fig:Spinarrangement}
\end{figure*}

In the AF phase, for $T \ll T_{\rm N}$, the ordered magnetic moments generate a local magnetic field $B_{\rm loc}$ at the stopping site of the muon, which is related to the sublattice magnetization of the Cu$^{2+}$ electronic moments. By using first-principles cluster calculations~\cite{Suter2003} the muon stopping site has been located at (0.119, 0.119, 0.2128) in the orthorhombic unit cell, $1.0$\,\AA{} off the apical oxygen, as in a oxygen-hydrogen bond (circles in Fig.~\ref{fig:Spinarrangement}). The ZF asymmetry spectra $A\,P^{\rm AF}_{\rm ZF}(t)$ can then be described by

\begin{equation}
A\,P^{\rm AF}_{\rm ZF}\left(t\right) = \sum_{i} A_{\rm T_{i}} \, \cos(\gamma_{\mu} \, B_{\rm loc,i} \, t + \phi) \, {\rm e}^{-\lambda_{\rm T_{i}} \, t} + A_{\rm L} \, {\rm e}^{-\lambda_{\rm L} \, t},
\label{eq:cos}
\end{equation}
where $A_{\rm T}$ and $A_{\rm L}$ reflect the fraction of the muons having their spin initially transverse and longitudinal to the internal field direction, respectively. The relaxation rate $\lambda_{\rm T}$ is proportional to the width of the internal field distribution sensed by the muon. In the presence of disorder $\lambda_{\rm T}$ can be larger than $\gamma_{\mu} B_{\rm loc}$, resulting in an overdamped asymmetry spectrum without oscillations. In the presence of fluctuating magnetic fields the longitudinal part of the muon spin polarization is relaxing as well with the corresponding rate $\lambda_{\rm L}$ ($\lambda_{\rm L} <0.1\,\mu$s$^{-1}$ likely due to nuclear dipole depolarization only). The phase $\phi$ is in general a temperature-independent constant.

\begin{figure}[t!]
	\centering
		\includegraphics[width=1\columnwidth]{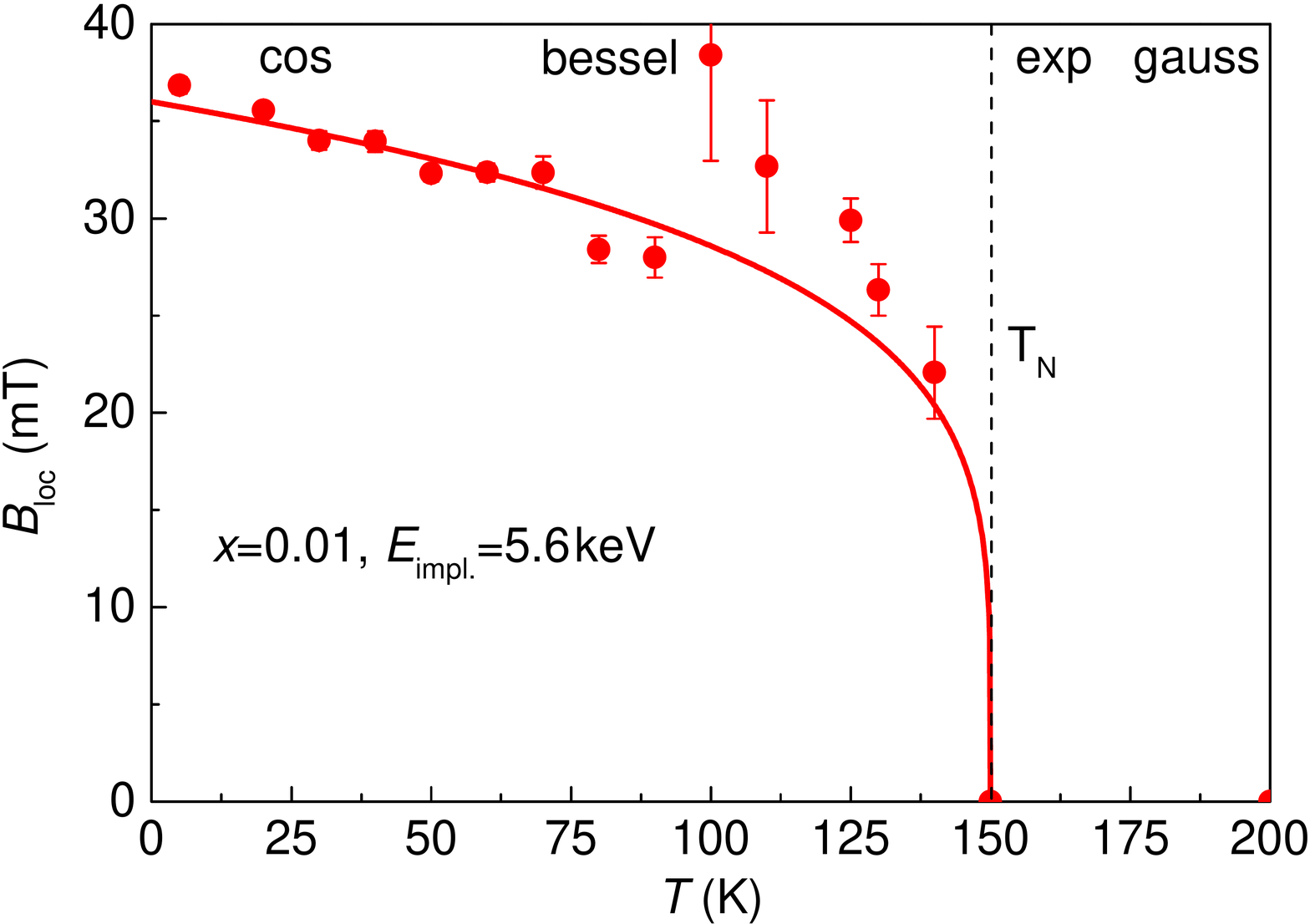}
	\caption{(color online) The temperature dependence of the local magnetic field $B_{\rm loc}$ at the muon stopping site for the LSCO film with $x=0.01$ for $E_{\rm impl.}=5.6$\,keV. The red solid line is a fit to the data using $B_{\rm loc}(T)=B_{\rm loc}(0{\rm K}) \cdot (1-T/T_{\rm N})^{0.21}$.}
	\label{fig:BversusT1}
\end{figure}
In the AF phase for $T \lesssim T_{\rm N}$, the strongly damped oscillations in $A\,P^{\rm AF}_{\rm ZF}(t)$ can be better described by a Bessel function of the first kind $J_0(t)$, which at larger times is equivalent to a cosine with a phase shift of~$\Delta \phi = 45^{\circ}$ and an additional damping of~$\sqrt{2/(\pi \, \gamma_{\mu} \, B_{\rm loc} \, t)}$, see Fig.~\ref{fig:LCO}\,(c)~and\,(g). When using the pure cosine function the phase $\phi$ strongly increases with increasing temperature, from less than $10^{\circ}$ at $5$\,K to more than $40^{\circ}$ at higher temperatures. In this case the Bessel function provides a better description of the measured $A\,P^{\rm AF}_{\rm ZF}(t)$:
\begin{equation}
\begin{split}
A\,P^{\rm AF,2}_{\rm ZF}\left(t\right)&= \, A_{\rm T} \, {\rm e}^{-\lambda_{\rm T} t} \, J_0\left(\gamma_{\mu} \, B_{\rm loc} \, t\right) \ + \, A_{\rm L} \, {\rm e}^{-\lambda_{\rm L} t}\\
 &\approx \, A_{\rm T} \, {\rm e}^{-\lambda_{\rm T} t} \, \sqrt{\displaystyle\frac{2}{\pi \, \gamma_{\mu} \, B_{\rm loc} \, t}} \, \cos\left(\gamma_{\mu} \, B_{\rm loc} \,t - \frac{\pi}{4}\right)\\
&\phantom{\approx} \, + \, A_{\rm L} \, {\rm e}^{-\lambda_{\rm L} t}.
\end{split}
\label{eq:AF}
\end{equation}

This behavior may arise from incommensurate magnetism~\cite{Yaouanc11}, where the period of the magnetic structure is not an integer multiple of the lattice constant, or from the presence of nanometer scale AF domains. Both cases lead to an asymmetric magnetic field distribution, which is better described by a Bessel function. In neutron diffraction studies on bulk material incommensurate magnetism was only observed for $x>0.05$ and $T<7$\,K~\cite{Yamada98}. For $T \ll T_{\rm N}$, $A\,P^{\rm AF}_{\rm ZF}(t)$ is well described by a cosine function as in bulk samples, see Fig.~\ref{fig:LCO}\,(d)~and\,(h).

In the AF state, the relative strength of the parameters $A_{\rm T}$ and $A_{\rm L}$ [see Eq.~(\ref{eq:cos})] reflects the local magnetic field distribution which is determined by the spatial arrangement of the Cu electronic magnetic moments. If the field at the muon stopping site is isotropic, corresponding to an electronic moment vector pointing with equal probability in all three directions, then the ratio $A_{\rm T}$:$A_{\rm L}$ is $\frac{2}{3}$:$\frac{1}{3}$. If the spins are aligned within the CuO$_2$ planes, corresponding to a field at the stopping site with only planar components, the ratio $A_{\rm T}$:$A_{\rm L}$ is $\frac{1}{2}$:$\frac{1}{2}$. Both LSCO samples ($x=0.00$ and $x=0.01$) show at the lowest temperature a ratio close to $\frac{1}{2}$:$\frac{1}{2}$. This result is in agreement with neutron data which revealed that the Cu electronic magnetic moments are preferentially aligned in the CuO$_2$ planes~\cite{Vaknin87}.

As evidenced by the beating in the asymmetry spectrum two frequencies, corresponding to two local fields, are observed in LCO thin films~[Fig.~\ref{fig:LCO}\,(d)]. Extrapolating the temperature dependence of the measured fields to $T=0$\,K by a power law yields $B_{\rm loc,1}(0\,\,{\rm K})=40.9(4)$\,mT and $B_{\rm loc,2}(0\,{\rm K})=11.2(1)$\,mT. For LCO bulk samples only one local magnetic field $B_{\rm loc}(0\,{\rm K}) \cong 43$\,mT has been reported~\cite{Borsa95}, although there are hints of a similar lower second field from unpublished data of high quality single crystals. A possible explanation for the appearance of an additional local magnetic field is a mixture of different alignments of the Cu electronic magnetic moments within the CuO$_2$ planes. From powder neutron diffraction experiments on LCO the spin structure shown in Fig.~\ref{fig:Spinarrangement}\,(a) was determined~\cite{Vaknin87}. The electronic spins of different CuO$_2$ planes (black and green) are aligned parallel or antiparallel to each other in the orthorhombic unit cell. By taking only dipole magnetic fields arising from the Cu electronic magnetic moments into account ($m_{\rm Cu}=0.645\, \mu_{\rm B}$~\cite{Pozzi99}), this arrangement leads to the magnetic field distribution in the plane of the muon stopping site shown in Fig.~\ref{fig:Spinarrangement}\,(e). Since the muon stops close to the apical oxygen, the same magnetic field value is present at crystallographically equivalent muon stopping sites [marked with circles in Fig.~\ref{fig:Spinarrangement}\,(e)]. Therefore, only \emph{one} $B_{\rm loc}$ is observable in this ``domain A''. A similar AF ordering but with rotated Cu spins within the CuO$_2$ planes, see for example Fig.~\ref{fig:Spinarrangement}~(b)~-~(d), generates two different magnetic field values at crystallographically equivalent muon stopping sites indicated by ``domain B'' [see Fig.~\ref{fig:Spinarrangement}\,(f)]. The calculated field values differ from the local magnetic fields determined by LE-$\mu$SR. This deviation is not surprising, because transferred hyperfine fields as well as higher-order corrections, \textit{e.g.} due to the Dzyaloshinskii-Moriya (DM) interactions, which could change the field values by a factor of two, are neglected here. To determine the magnetic field values precisely full density functional theory calculations have to be performed. The asymmetry $A_{\rm T,1}=0.074$ (related to $B_{\rm loc,1}$) is more than two times larger compared to $A_{\rm T,2}=0.030$ (related to $B_{\rm loc,2}$). Since the ZF oscillation amplitudes $A_i$ are proportional to the magnetic volume fractions of the domains in the sample a mixture of two spin arrangements has to be present. While $B_{\rm loc,1}$ is present in both domains but $B_{\rm loc,2}$ only in domain B, the asymmetry ratio corresponds to a volume ratio domain A to domain B of $42\,\%$:$58\,\%$ for the two spin arrangements.
\begin{figure}[t!]
	\centering
		\includegraphics[width=1\columnwidth]{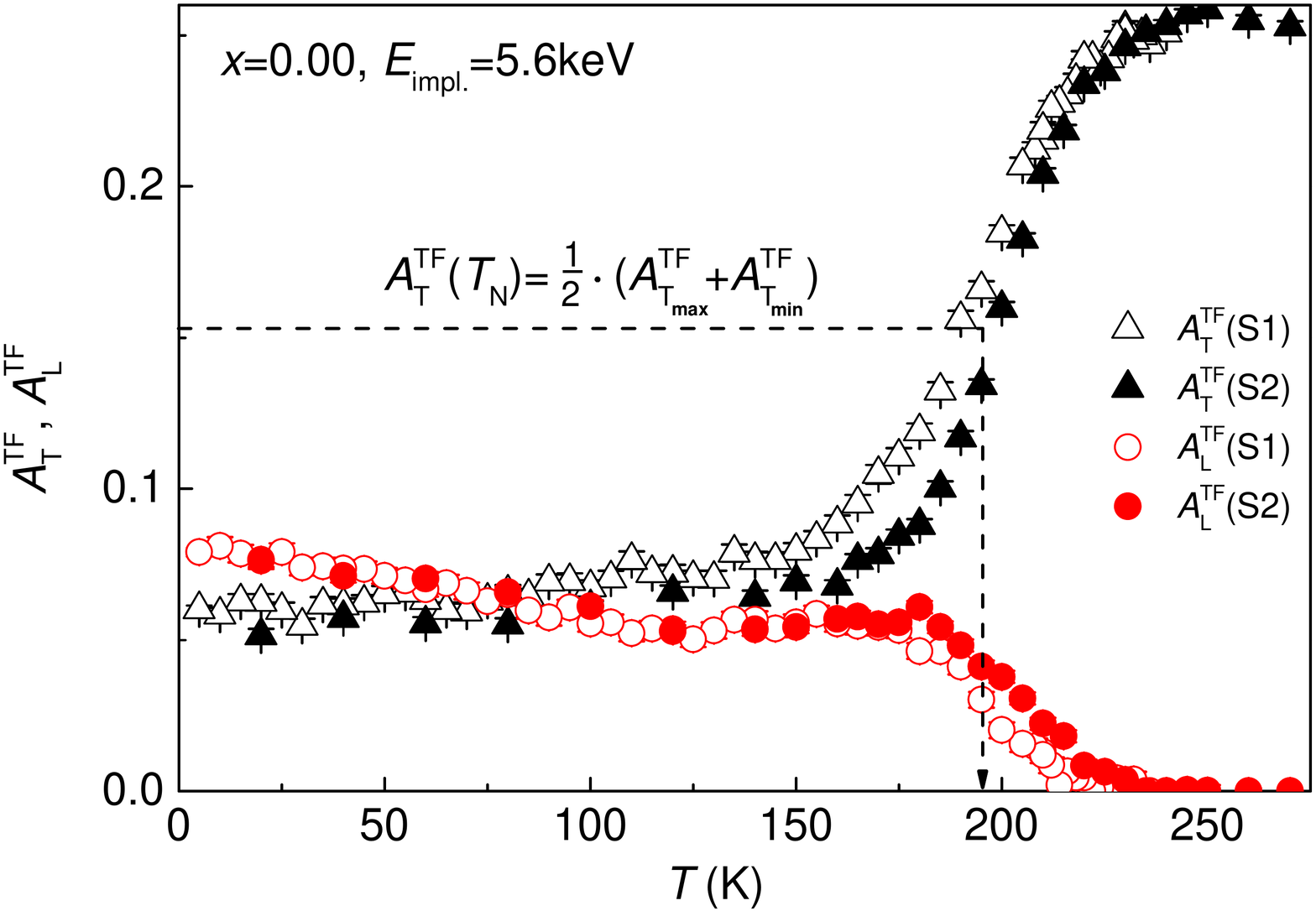}
		\caption{(color online) The transverse ($A^{\rm TF}_{\rm T}$) and longitudinal ($A^{\rm TF}_{\rm L}$) asymmetry as a function of $T$ determined in an external magnetic field of $B_{\rm ext}=2.9$\,mT for $E_{\rm impl.}=5.6$\,keV for two sets of LSCO thin film samples (S1,S2) with $x=0.00$. The dashed arrow shows how $T_{\rm N}$ was determined.}
	\label{fig:Asym}
\end{figure}

Different Cu spin arrangements could originate from structural changes. In general, the $c$ axis of thin-film samples grown on a LSAO substrate is larger compared to the bulk value. These changes in the crystal structure have a strong influence on the anisotropic parts of the spin Hamiltonian~\cite{Coffey91}, and hence on the spin configuration. This is consistent with observations for LCO crystallizing in the metastable tetragonal Nd$_2$CuO$_4$ structure with a $c$ axis lattice parameter of $12.52$\,nm only~\cite{Hord10}. Bulk $\mu$SR~\cite{Hord10} revealed a lower internal magnetic field of $B_{\rm loc}(0\,{\rm K})=11$\,mT and a different spin arrangement is expected compared to LCO in the orthorhombic phase. In our study no changes are observed in the $c$ axis lattice parameter compared to bulk values. So the existence of a mixture of tetragonal and orthorhombic phases seems unlikely. It is more likely that magnetic domains with different spin arrangements are present. This would also explain that in some bulk samples a second field is observed, where no tetragonal structure exists and $T_{\rm N}$ is higher compared to thin films. A change in the Cu spin arrangement could be caused by dislocations or defects, which could arise in thin films from the lattice mismatch between LSCO and LSAO.

For the nominal Sr content $x=0.01$ the generic behavior of $A P_{\rm ZF}(t)$ as function of $T$ is similar as for $x=0.00$, but only one $B_{\rm loc}$ is present just like in bulk samples [Fig.~\ref{fig:LCO}\,(h)]. In bulk samples with $x=0.01$ spin freezing of the doped holes is observed below $T_{\rm f}=8$\,K (Fig.~\ref{fig:Phasediagram}). As determined by ZF $\mu$SR measurements, this freezing below $T_{\rm f}$ manifests itself as an drastic increase of the slope ${\rm d}B_{\rm loc}/{\rm d}T$ (see Ref.~\onlinecite{Borsa95}). In the present study no increase is observed down to $5$\,K (Fig.~\ref{fig:BversusT1}). Thus the shape of $B_{\rm loc}(T)$ and the fact that $B^{x=0.01}_{\rm loc}(0\,{\rm K})=36.0(5)$\,mT is below $B^{x=0.00}_{\rm loc}(0\,{\rm K})=40.9(4)$\,mT indicate a strong suppression of hole spin freezing in LSCO thin film samples.

\begin{figure}[t!]
	\centering
		\includegraphics[width=1\columnwidth]{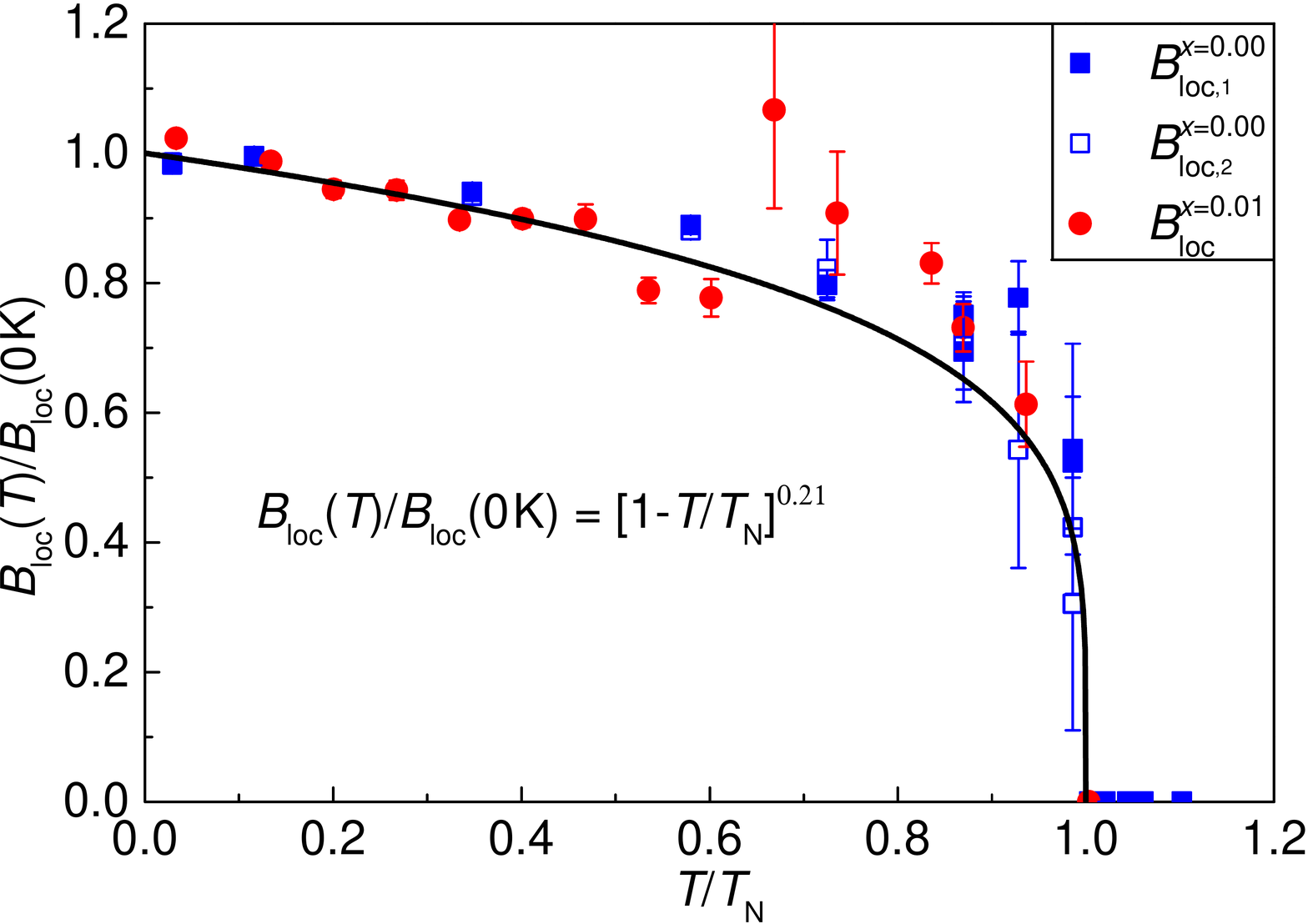}
	\caption{(color online) The normalized internal magnetic field $B_{\rm loc}(T)/B_{\rm loc}(0\,{\rm K})$ as a function of the normalized temperature $T/T_{\rm N}$ for LSCO thin films ($x=0.00$ and $x=0.01$). The solid black line corresponds to the power law given in Eq.~(\ref{eq:BofT}) with $\beta = 0.21$ and $B^{x=0.00}_{\rm loc,1}(0\,{\rm K})=41$\,mT, $B^{x=0.00}_{\rm loc,2}(0\,{\rm K})=11$\,mT, and $B^{x=0.01}_{\rm loc}(0\,{\rm K})=36$\,mT.}
	\label{fig:BversusT}
\end{figure}

The N\'{e}el temperatures as well as the magnetic volume fractions were determined from temperature scans in a weak magnetic field $\mathbf{B}_{\rm ext}$ applied perpendicular to the initial muon spin polarization and to the film surface. The asymmetry time spectra $A\,P_{\rm wTF}(t)$ in a wTF are described by:
\begin{equation}
\label{eq:wTF}
\begin{split}
  A P_{\rm wTF}(t) =& \quad \, A^{\rm TF}_{\rm T} \, \cos(\gamma_\mu B_{\rm ext} t + \phi) \, {\rm e}^{-\frac{1}{2} \sigma_{\rm T}^2 t^2} \\
                    &+  A^{\rm TF}_{\rm L} \,  \cos(\phi) \, {\rm e}^{-\lambda_{\rm L} t}.
\end{split}
\end{equation}
Eq.~\ref{eq:wTF} represents the parramagnetic part of the muon spin polarization. The superposition of the antiferromagnetic and the applied field leads to a strong damping of about $\gtrsim 50 \mu$s$^{-1}$ of the full polarization which has been neglected in the fit. $A^{\rm TF}_{\rm T}$ and $A^{\rm TF}_{\rm L}$ are the transverse and longitudinal oscillation amplitudes. Above $T_{\rm N}$, $A^{\rm TF}_{\rm T}$ is the full asymmetry, since only $\textbf{B}_{\rm ext}$ is present. Below $T_{\rm N}$, the superposition of the small external and the internal magnetic fields leads to a strong dephasing of the signal, so $A^{\rm TF}_{\rm T}$ decreases to a level corresponding to the non-magnetic fraction plus the background level. $A^{\rm TF}_{\rm L}$ represents the part of non-precessing muon spins. A decrease in $A^{\rm TF}_{\rm T}$ with a simultaneous increase in $A^{\rm TF}_{\rm L}$ demonstrates static magnetism. $\sigma_{\rm T}$ is the depolarization rate of the precessing muon fraction and reflects the field width observed by the muons in the nonmagnetic parts of the sample. At the lowest temperature it is dominated by the nuclear magnetic moments of La and Cu. The depolarization rate $\lambda_{\rm L} \simeq 0$ for all measurements, and $\phi$ is the temperature-independent detector phase.

\begin{figure}[t!]
	\centering
		\includegraphics[width=1\columnwidth]{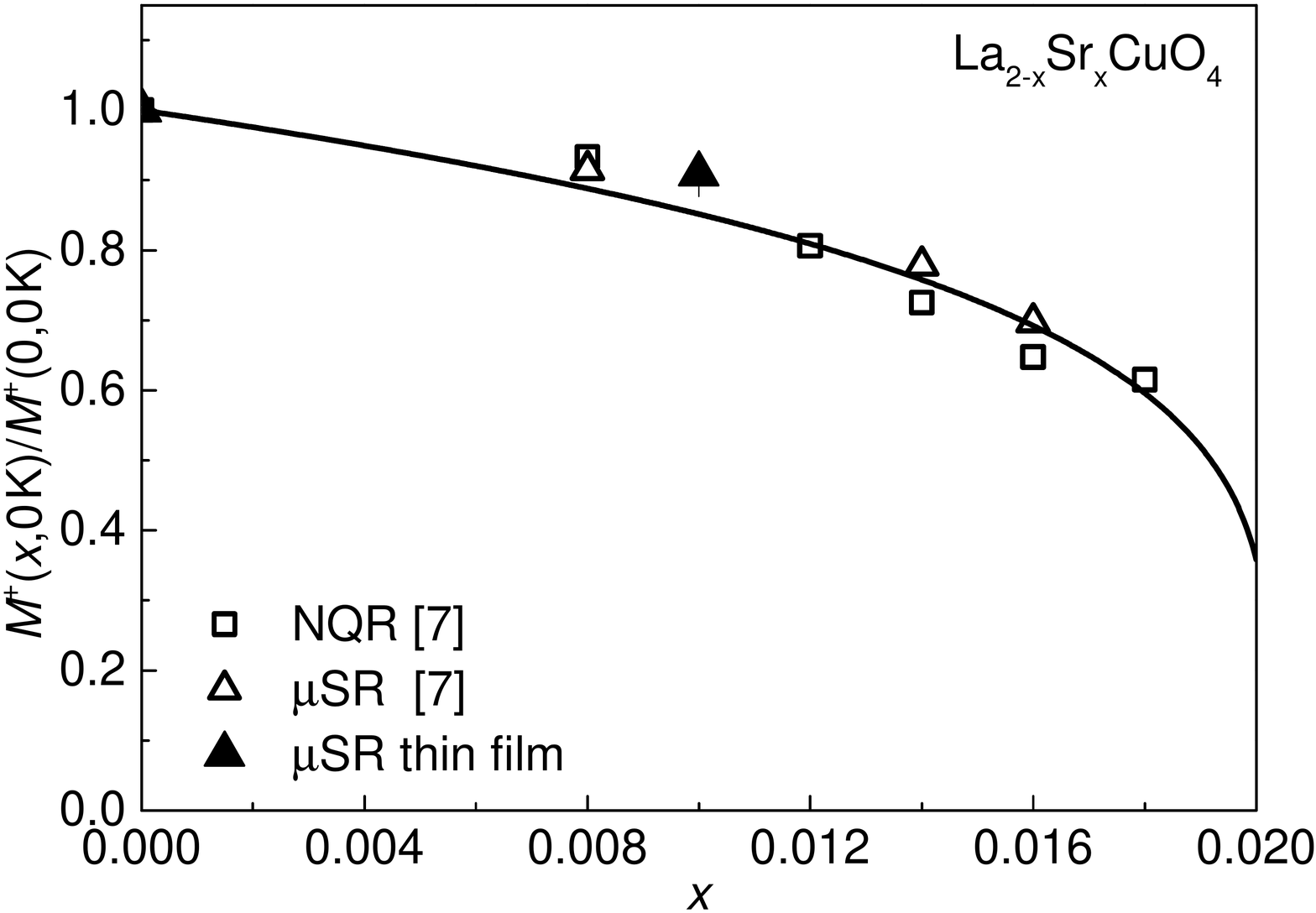}
	\caption{The normalized staggered magnetization $M^{+}(x,0\,{\rm K})/M^{+}(0,0\,{\rm K})$ as a function of the nominal Sr content $x$ of LSCO thin films (solid triangles) and LSCO bulk samples as inferred from NQR and $\mu$SR experiments (open symbols) given in Ref.~\onlinecite{Borsa95}. The black line is a fit to the bulk data using Eq.~(\ref{eq:MofX}).}
	\label{fig:BversusX}
\end{figure}

\begin{figure*}[t]
\centering
    \includegraphics[width=1.95\columnwidth]{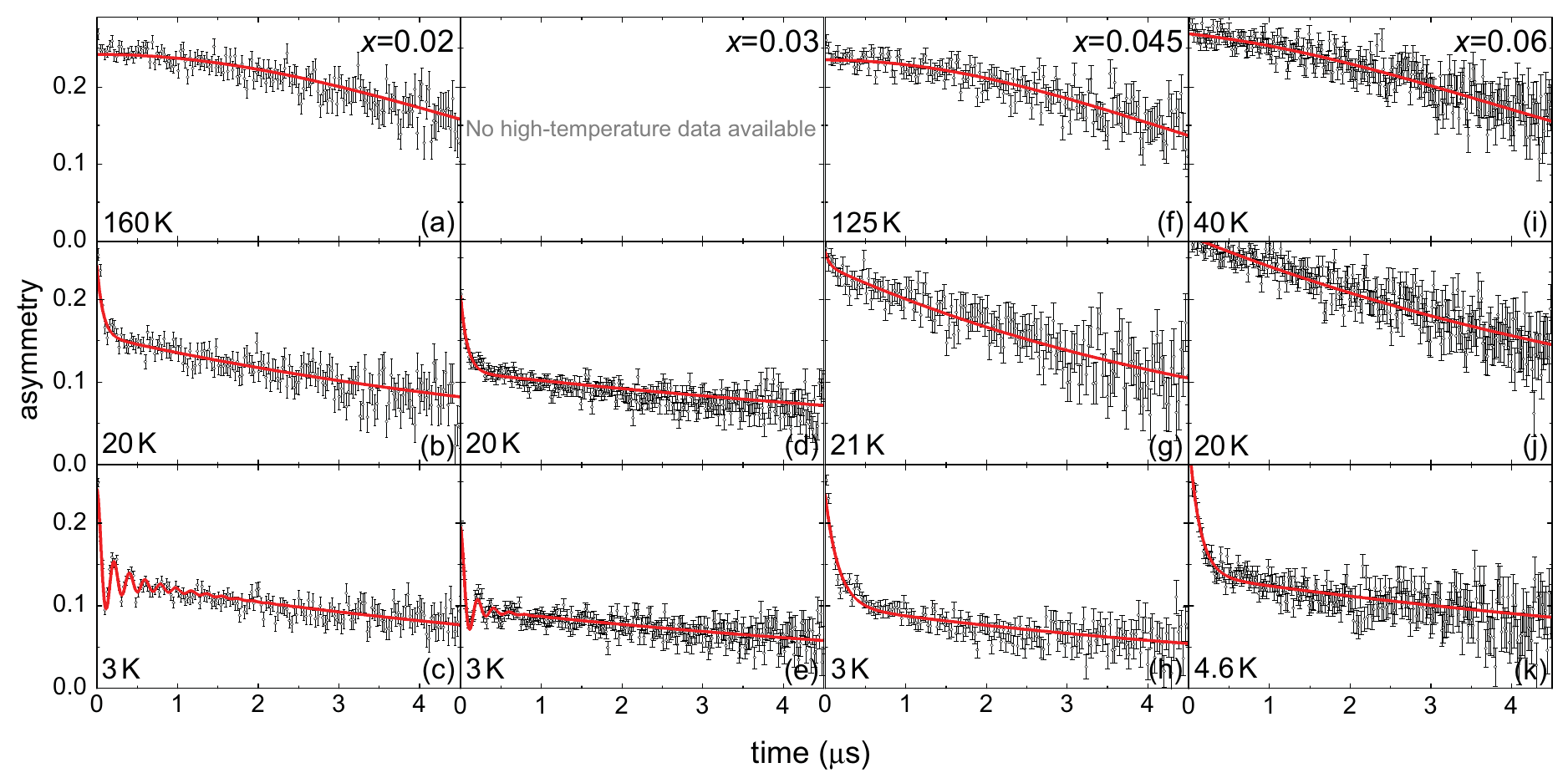}  
  \caption{(color online) The ZF asymmetry time spectra of LSCO thin films with $x=0.02, 0.03, 0.045,$ and $0.06$ at different temperatures as indicated for $E_{\rm impl.}=5.6$\,keV. The solid red lines are fits to the data done with musrfit~\cite{Suter12}. See text for more details.}
  \label{fig:LSCO2}
\end{figure*}

The magnetic volume fraction $f$ is given by
\begin{equation}
f = \frac{A^{\rm TF}_{\rm T_{\rm max}}-A^{\rm TF}_{\rm T_{\rm min}}}{A^{\rm TF}_{\rm T_{\rm max}}-A_{\rm Bkg}},
\label{eq:f}
\end{equation} 
taking into account a constant background asymmetry of $A_{\rm Bkg}=0.17(3) \cdot A^{\rm TF}_{\rm T_{\rm max}}$ as in ZF measurements. The determined volume fractions are listed in Table~\ref{tab:Tg}. The magnetic transition temperatures $T_{\rm N,g}$ were defined as the temperature for which
\begin{equation} 
 A^{\rm TF}_{\rm T}(T_{\rm N,f,g}) = \frac{1}{2} \cdot(A^{\rm TF}_{\rm T_{\rm max}}+A^{\rm TF}_{\rm T_{\rm min}}),
\label{eq:TNfg}
\end{equation}
yielding $T_{\rm N}^{x=0.00}=195(3)$\,K (see Fig.~\ref{fig:Asym}) and $T_{\rm N}^{x=0.01}=151(5)$\,K. Both values are well below the respective bulk values of $T_{\rm N}^{x=0.00}\simeq300$\,K and $T_{\rm N}^{x=0.01} \simeq 250$\,K (see Fig.~\ref{fig:Phasediagram}). The relation between the normalized internal magnetic field $B_{\rm loc}(T)/B_{\rm loc}(0\,{\rm K})$ and the normalized temperature $T/T_{\rm N}$ can be analyzed using~\cite{Borsa95}:
\begin{equation}
\frac{B_{\rm loc}\left(T\right)}{B_{\rm loc}\left(0\,{\rm K}\right)}=\left[1-\frac{T}{T_{\rm N}}\right]^{\beta}.
\label{eq:BofT}
\end{equation}

The obtained exponent $\beta$ is similar for both Sr contents $x$ as well as for $B_{\rm loc,1}$ and $B_{\rm loc,2}$, suggesting a common underlying ordering mechanism. The thin-film data are well described with $\beta=0.21$ found in the bulk~\cite{Borsa95} (Fig.~\ref{fig:BversusT}). Furthermore, the doping dependence of the normalized staggered magnetization $M^{+}(x,0\,{\rm K})/M^{+}(0,0\,{\rm K})\propto B_{\rm loc}(x,0\,{\rm K})/B_{\rm loc}(0,0\,{\rm K})$ in the thin-film samples of the present work are in agreement with nuclear quadrupole resonance (NQR) and $\mu$SR results obtained for bulk samples~\cite{Borsa95} (Fig.~\ref{fig:BversusX}). The staggered magnetization follows the empirical relation given in Ref.~\onlinecite{Borsa95}:
\begin{equation}
M^{+}(x,0\,{\rm K})/M^{+}(0,0\,{\rm K}) = \left[1-\frac{x}{x_{\rm c}}\right]^{n},
\label{eq:MofX}
\end{equation}
with a critical doping of $x_{\rm c}=0.0203$ and an exponent $n=0.236$ (Fig.~\ref{fig:BversusX}).

In summary, whereas in the AF phase of LSCO the generalized behavior of $B(T)/B(0\,{\rm K})$ as a function of $x$ and $T/T_{\rm N}$ is similar in thin films and bulk samples (Figs.~\ref{fig:BversusT}~and~\ref{fig:BversusX}), $T_{\rm N}$ is strongly suppressed, and $T_{\rm f}$ is not observed down to $5$\,K in thin films. The local magnetic fields at the muon stopping site are instead very similar in bulk and thin films, indicating an equal magnitude of ordered electronic moments.

\subsection{Disordered Magnetic Phase for $x \gtrsim 0.02$}
\label{sec:DisorderedMP}

In LSCO films with $x=0.02$ and $x=0.03$ the ZF asymmetry time spectra show an oscillation at the lowest temperatures [Fig.~\ref{fig:LSCO2}\,(c) and~(e)], while the temperature dependence of $B_{\rm loc}$ given by Eq.~(\ref{eq:BofT}) is changed drastically. The best fit to $B_{\rm loc}(T)$ for $x=0.02$ yields $\beta=0.04(2)$. Since $\beta \simeq 0.2$ is characteristic for the AF phase, samples with $x \gtrsim 0.02$ have to be instead in the peculiar low temperature magnetic phase (Fig.~\ref{fig:Phasediagram}), leading to ZF $\mu$SR precession too. This phase is termed in the literature as ``spin-glass'' or ``cluster spin-glass'' (CSG) phase~\cite{Chou93,Niedermayer98,Rigamonti06}. Although somewhat misleading, we adopt this terminology for consistency with the literature. In the CSG phase dynamical spin and charge stripes have been found in some cuprate systems~\cite{Tranquada95,Bianconi96}. Microsegregation of mobile holes leads to hole-poor AF areas separated by hole-rich nonmagnetic domain walls. The presence of charge or spin density waves within the CSG phase is another proposed state~\cite{Torchinsky2013}. At low temperatures the dynamics of the CSG state slow down and oscillations are observed in the ZF asymmetry time spectra in bulk samples~\cite{Niedermayer98}. For LSCO thin-film samples oscillations are observed at low temperatures too [Fig.~\ref{fig:LSCO2}\,(c)~and~(e)]. At $3$\,K the Bessel function [Eq.~(\ref{eq:AF})] describes the obtained $\mu$SR data for $x=0.02$ and $x=0.03$ very well. This suggests the presence of incommensurate magnetism in the CSG phase as observed by neutron diffraction in bulk samples~\cite{Yamada98}.

\begin{table}[b]
\caption{Values of the transition temperatures $T_{\rm N,g}$ and the corresponding magnetic volume fractions $f$ for various nominal Sr contents $x$ of LSCO thin films obtained from wTF $\mu$SR data.}
\label{tab:Tg}
		\begin{ruledtabular}
				\begin{tabular}{lcc}
 $x$			& $T_{\rm N,g}$(K) 	& $f$(\%)	\\ \hline
 $0.00$		& $195(3)$					& $90(3)$	\\
 $0.01$ 	& $151(5)$					& $89(3)$	\\
 $0.02$		& $37(7)$						& $93(3)$	\\
 $0.03$ 	& $25(2)$						& $80(3)$	\\
 $0.045$	& $9(2)$       			& $93(3)$ \\
 $0.06$ 	& $7(1)$						& $93(3)$ \\
				\end{tabular}
		\end{ruledtabular}
\end{table}

For $x=0.045$ and $x=0.06$ the whole temperature scale is shifted down since $T_{\rm g} \propto 1/x$. Therefore no ZF oscillations were observed down to $3$\,K [Fig.~\ref{fig:LSCO2}\,(h)~and~(k)]. The asymmetry time spectra at $3$\,K ($x=0.045$) and $4.6$\,K ($x=0.06$) show instead a strong double-exponential behavior (sum of two exponential functions) with considerably enhanced depolarization rates, as present at $20$\,K for $x=0.02$ and $x=0.03$. For $T \gtrsim T_{\rm g}$ all asymmetry time spectra for $x\geq0.02$ show an exponential decay [Fig.~\ref{fig:LSCO2}\,(g) and~(j)], while for $T \gg T_{\rm g}$ they are described by a Gaussian Kubo-Toyabe function [Fig.~\ref{fig:LSCO2}\,(a),\,(f), and~(i)]. Therefore, the LSCO $\mu$SR spectra for all investigated samples show the same behavior, namely a slowing down of electronic fluctuations in the PM phase.

From wTF $\mu$SR measurements the glass transition temperature $T_{\rm g}$ and the magnetic volume fraction $f_{\rm CSG}$ were determined for different $x$ with the method described in the previous section [Eqs.~(\ref{eq:f}) and~(\ref{eq:TNfg})]. The corresponding values are listed in Table~\ref{tab:Tg}. The present values of $T_{\rm g}$ are all significantly larger than those determined by bulk $\mu$SR~\cite{Niedermayer98,Borsa95} (see Fig.~\ref{fig:Phasediagram}). This difference could not be ascribed to the method for determining $T_{\rm g}$. If we define $T_{\rm g}$ in the same way as in Refs.~\onlinecite{Borsa95} and~\onlinecite{Niedermayer98} we obtain consistent values.

\subsection{Discussion}
\label{sec:Discussion}

What is the origin of the differences between the thin-film and the bulk phase diagrams? There are various potential mechanisms and parameters which may modify the phase diagram, such as oxygen off-stoichiometry, strain, geometric frustration or defects.
 
\begin{figure}[t!]
	\centering
		\includegraphics[width=\columnwidth]{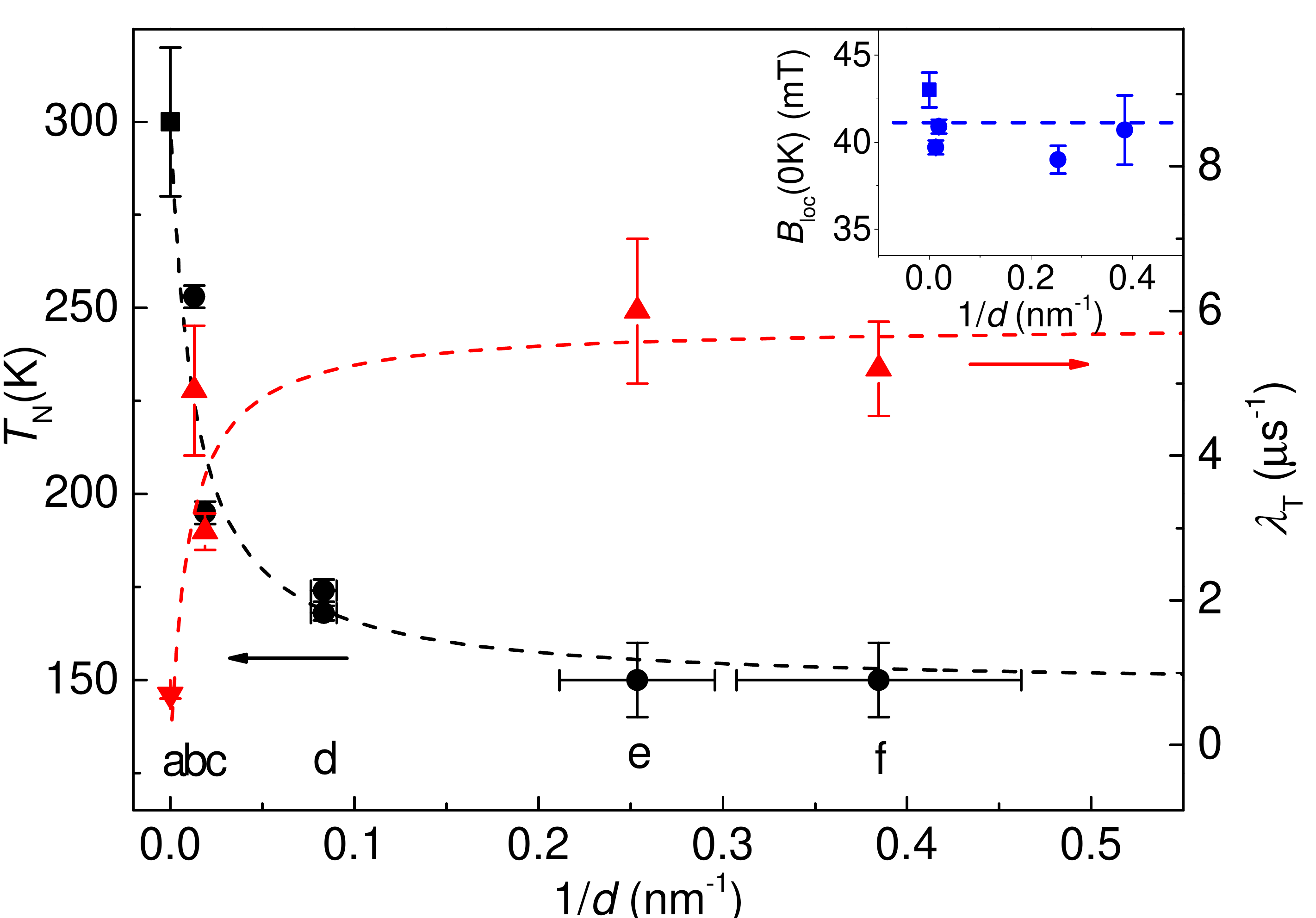}
	\caption{(color online) The N\'{e}el temperature $T_{\rm N}$, the ZF depolarization rate $\lambda_{\rm T}$ and the internal magnetic field $B_{\rm loc}(0\,{\rm K})$ (inset) as function of the inverse sample thickness $1/d$ for different LCO samples. The bulk thickness is set to infinity. The black, red, and blue dashed lines are guides to the eye. a: LCO bulk material~\cite{Niedermayer98,Kastner98,Cho93,Chou93,Borsa95}; b: LCO/STO thin film (unpublished); c: present data; d: LCO/STO and LCO/LSAO thin films~\cite{Suter04}; e: LCO/LSCO superlattices~\cite{Suter2011}; f: LCO/LaAlO$_{3}$ superlattices (unpublished).}
	\label{fig:Thickness}
\end{figure}

Oxygen off-stoichiometry would indeed lead to an increase of doped holes in the CuO$_2$ planes and hence to a $x_{\rm eff} > x$. This in turn would yield a lower $T_{\rm N}$ (Ref.~\onlinecite{Wells97}) as found in this study (Fig.~\ref{fig:Phasediagram}). However, at the same time $T_{\rm f}$ should increase, while $T_{\rm g}$ should decrease. This is the exact opposite to our observation. The case $x_{\rm eff} > x$ is also unlikely, since $B_{\rm loc}$ is the same in bulk material and thin films (Figs.~\ref{fig:BversusX} and~\ref{fig:Thickness}). Furthermore, the border between the AF and the CSG phase ($x \simeq 0.02$) is not shifted (Fig.~\ref{fig:Phasediagram}). Additional interstitial oxygen in the films would expand the $c$ axis lattice parameter~\cite{Butko09} different to our observations: $c^{x=0.00}=13.15(2)\, {\rm \AA{}} =c^{\rm bulk}$ (Fig.~\ref{fig:Xray}). In order to double-check the possible effects of the variations in oxygen stoichiometry, we post-annealed two sets of LCO films in high vacuum, using two \emph{different} procedures (temperature and time). Both sets show the same $T_{\rm N}$ and the same temperature dependence of the staggered magnetization. Therefore, oxygen off-stoichiometry is quite small and cannot be the dominant source of the differences between the bulk and the thin-film samples.

For undoped cuprates $T_{\rm N}$ is related to the inter-plane coupling constant $J'$ and the 2D in-plane correlation length $\xi_{\rm 2D}$ by
\begin{equation}
k_{\rm B} T_{\rm N} \simeq J' \cdot (m^{+})^2 \cdot \left[\frac{\xi_{\rm 2D}(J,T_{\rm N},y)}{\alpha}\right]^2,
\end{equation} 
where $\alpha$ is the distance between the copper moments and $m^{+}$ the reduced magnetic moment~\cite{Chakravarty88,Chakravarty90,Hasenfratz91}. $J'$ might be sensitive to strain, whereas $\xi_{\rm 2D}$ is influenced by the in-plane coupling constant $J$ (Ref.~\onlinecite{Aronson91}) and the amount of disorder $y$ (Ref.~\onlinecite{Chen00}). A reduction of  $J'$ and/or $\xi_{\rm 2D}$ would lead to the observed decrease in $T_{\rm N}$.

In the following possible strain effects will be discussed, assuming the absence of any disorder ($y=0$). In this case $\xi_{\rm 2D}$ is given by
\begin{equation}
\frac{\xi_{\rm 2D}\left(J,T_{\rm N},0\right)}{\alpha} = 0.567 \cdot \frac{J}{2 \pi \rho_{\rm s}} \cdot {\rm e}^{\frac{2 \pi \rho_{\rm s}}{k_{\rm B} T_{\rm N}}} \left[1- \frac{k_{\rm B} T_{\rm N}}{4 \pi \rho_{\rm s}}\right],
\end{equation}
with a spin stiffness $2 \pi \rho_{\rm s}=0.94\,J$ for LCO~\cite{Hasenfratz91,Aronson91}.

Poisson strain is likely to modify $J$ as well as $J'$, because the lattice parameters are changed keeping the unit cell volume constant. When compared to its bulk value, the $c$~axis is in general enlarged for LSCO grown on LSAO due to compressive strain of the substrate~\cite{Tsukada02} (as long as the sample thickness is below the critical value of about $20$ unit cells~\cite{Butko09}), reducing $J'$. At the same time LCO grown on LSAO should exhibit a higher $J$ because of the changed in-plane lattice constants [$J \propto 1/\alpha^{6.4}$ (Ref.~\onlinecite{Aronson91})]. To reach the observed $T_{\rm N}$, a $J'_{\rm film} \approx 10^{-2} J'_{\rm bulk}$ is required. Since the $c$~axis lattice constants of the films are very close to the bulk values, such a strong reduction of $J'$ is unlikely. Thus, Poisson strain is not the main reason for the drastic $T_{\rm N}$ reduction.

In strain released LCO thin films the $a$ and $b$ lattice parameters differ from their bulk values~\cite{Butko09} through Madelung strain, leading to a smaller unit cell volume as also obtained by applying hydrostatic pressure. Raman scattering studies on AF single crystal LCO~\cite{Aronson91} showed that pressure leads to an enhancement of $J$ and therefore to an increase of $T_{\rm N}$.  Therefore, $J$ and $T_{\rm N}$ should be also increased in LCO thin films. This is the opposite to our observations. A reduction of the N\'eel temperature through Madelung strain is therefore unlikely.

Geometrical frustration within a system may also influence the transition temperatures. A low asymmetry between the in-plane lattice constants $r=1-a/b$ could lead to a reduced $J'$. A tetragonal system ($r=0$) consists of perfectly geometrically frustrated Cu electronic moments, since the CuO$_2$ layers within one unit cell are shifted by half a unit cell against each other [Fig.~\ref{fig:Spinarrangement}\,(a)]. A more orthorhombic system ($r \neq 0$) is less frustrated and exhibits a larger $J'$. In LCO thin films a lower in-plane lattice constants asymmetry is observed compared to bulk values~\cite{Butko09}: $r_{\rm film}=0.001 < r_{\rm bulk}=0.01$. Thin films are hence more frustrated which leads to a reduction of $J'$, resulting in a lower $T_{\rm N}$. But if geometrical frustration would be the main source of the reduced N\'eel temperature a similar system with $r=0$  should have an even lower $T_{\rm N}$ because of a lower $J'$. Sr$_2$CuCl$_2$O$_2$ (SCCO) is such a system. It exhibits almost the same in-plane coupling constant ($J_{\rm SCCO}/k_{\rm B} = 1450\,{\rm K} \approx J_{\rm LCO}/k_{\rm B}$), but at the same time a reduced Cu electronic magnetic moment (SCCO: $m _{\rm Cu} = 0.31 \mu_{\rm B}$~\cite{Greven94}, LCO: $m _{\rm Cu} = 0.645 \mu_{\rm B}$~\cite{Pozzi99}). Although SCCO is perfectly frustrated and $J'_{\rm SCCO}~\simeq~10^{-6}\,J~<~J'_{\rm LCO}~\simeq~10^{-5}\,J $ it shows a $T_{\rm N}=256$\,K~\cite{Greven94} which is well above that one observed for LCO thin films ($T_{\rm N}=195$\,K, $d=53$\,nm). Even though geometrical frustration will lead to a reduction of $T_{\rm N}$, the observed reduction is too substantial to originate from this source only.

What might influence the magnetic ground state as well are higher-order terms which are present in addition to the dominant super-exchange, like next-nearest-neighbor exchange. It has been shown that especially the DM interaction is very sensitive to the crystal symmetry~\cite{Coffey91}, which could naturally explain the spin re-orientation discussed in Sec.~\ref{sec:AF}. However, it is unlikely that these higher-order corrections will have a substantial effect on $T_{\rm N}$, $T_{\rm f}$, or $T_{\rm g}$ as observed in this study.

Epitaxial thin films differ from the bulk samples by the presence of strain-induced defects, such as stacking faults and misfit dislocations. The latter have been observed in high-resolution cross-section transmission electron microscopy measurements in LCO films~\cite{He07}. Typically, the defect density in stress released thin films is much higher compared to bulk samples. Depending on the nature of the defect, it can lead to charge trapping or pinning of collective modes like charge stripes, charge density waves, or spin density waves (weak collective pinning in the case of point defects, or strong pinning in the case of dislocations), likely to be present at higher doping ($x \gtrsim 0.02$). This could indeed give rise to an increase of $T_{\rm g}$ as discussed by Shengelaya \textit{et al.} (Ref.~\onlinecite{Shengelaya99}). This picture is also supported theoretically as discussed in Refs.~\onlinecite{Chen00} and~\onlinecite{Murthy88} where the influence of short-length-quantum and long-range disorder on the spin-$1/2$ quasi-2D Heisenberg antiferromagnet on a square lattice (QHAF) is discussed. In Ref.~\onlinecite{Chen00} disorder by dilution is studied leading to an explicit expression for the reduction of $T_{\rm N}$ as function of dilution, which was experimentally verified by Carretta \textit{et al.} (Ref.~\onlinecite{Carretta95}). This dilution is likely to introduce also magnetic frustration which has been clarified by new experimentally results by Carretta \textit{et al.} (Ref.~\onlinecite{Carretta11}) and theoretically by Liu and Chernyshev (Ref.~\onlinecite{Liu13}). Whereas disorder and/or frustration by dilution are directly applicable to Zn and Mg doping in LSCO, it is probably not the case for thin films for which misfit dislocations are the most likely source of disorder. Murthy~\cite{Murthy88} showed that a QHAF is much more sensitive to random fields than to moderate  random in-plane couplings. Hence, misfit dislocations due to strain and strain release would have a much stronger influence on $T_{\rm N}$ as suggested in Ref.~\onlinecite{Chen00}. Unfortunately, no quantitative expression for the reduction of $T_{\rm N}$ has been derived in Ref.~\onlinecite{Murthy88}. The measured ZF depolarization rates $\lambda_{\rm T}$  [Eq.~(\ref{eq:cos})] support this interpretation. In Fig.~\ref{fig:Thickness} the N\'{e}el temperature~$T_{\rm N}$, the ZF depolarization rate $\lambda_{\rm T}$ as well as the local magnetic field~$B_{\rm loc}$($0$\,K) (inset) of LCO and LCO superlattices are plotted as a function of the inverse thickness~$1/d$. While $B_{\rm loc}$($0$\,K) stays constant, $T_{\rm N}$ decreases with decreasing thickness $d$ systematically. At the same time $\lambda_{\rm T}$ increases with decreasing $d$. The ZF depolarization rate $\lambda_{\rm T}$ is a measure of the magnetic disorder, which is related to the before mentioned random fields~\cite{Murthy88}. According to theory~\cite{Chen00,Murthy88} disorder leads to a reduction of $T_{\rm N}$ in agreement with Fig.~\ref{fig:Thickness}. So disorder seems to be a probable mechanism which could explain consistently the differences in $T_{\rm N}$ and $T_{\rm g}$ of bulk and thin-film magnetic phase diagrams.

Extended LE-$\mu$SR studies of thin films with different thicknesses on the same substrate and thin films with the same thickness on different substrates, would be necessary to test the presented interpretations.

\section{Summary and Conclusions}
\label{sec:conclusions}

In this study we determined the magnetic phase diagram of LSCO thin films (thickness $53$\,nm) in the doping range $0 \leq x \leq 0.06$. The absolute scales of the transition temperatures differ substantially between the bulk and thin film samples. The N\'{e}el temperatures $T_{\rm N}$ are strongly reduced in the thin films and in the AF region no spin freezing is observed down to $5$\,K. The CSG transition temperatures $T_{\rm g}$ lie well above the corresponding bulk values. Oxygen off-stoichiometry and strain-induced changes of the lattice parameters or higher-order magnetic coupling constants are unlikely to explain the observed differences. Misfit dislocations through strain release might well be at the heart of the discovered effects. Overall, the thin-film and bulk samples exhibit similar magnitude, temperature, and doping dependence of the staggered magnetization and the same border between the AF and the CSG phase ($x \simeq 0.02$). The determined magnetic phase diagram provides a solid basis for future studies of multilayer and superlattice LSCO thin films.

\begin{acknowledgments}
We gratefully acknowledge Hans-Peter Weber for his excellent technical support. This work was partly supported by the Swiss National Science Foundation. Research at Brookhaven was supported by the U.S. Department of Energy, Basic Energy Sciences, Materials Sciences and Engineering Division.
\end{acknowledgments}

\end{document}